\documentclass[twocolumn,tightenlines,prd,floatfix,showpacs,superscriptaddress]{revtex4}
\usepackage{epsfig}
\usepackage{amssymb}
\usepackage{amsmath}
\usepackage{amsfonts}
\usepackage{color}

\begin{document}

\title{Probing dark matter models with neutrinos from the Galactic center}

\author{Arif Emre Erkoca}
\affiliation{Department of Physics, University of Arizona, Tucson, AZ 85721}

\author{Mary Hall Reno}
\affiliation{Department of Physics 
and Astronomy, University of Iowa, Iowa City, IA 52242}

\author{Ina Sarcevic}
\affiliation{Department of Physics, University of Arizona, Tucson, AZ 85721}
\affiliation{Department of Astronomy and Steward Observatory,
University of Arizona, Tucson, AZ 85721}

\begin{abstract}

We calculate the contained and upward muon and 
shower fluxes due to neutrinos produced via 
dark matter annihilation or decay 
in the Galactic center. 
We consider dark matter models in which the dark matter particle is a gravitino, a Kaluza-Klein 
particle and a particle in leptophilic models. 
The Navarro-Frenk-White profile for the dark matter density distribution in the Galaxy is used. 
We incorporate 
neutrino oscillations by assuming maximal mixing and parametrize our results for muon
and shower distributions. The muon and shower event rates and the minimum observation times
in order 
to reach 
2$\sigma$ detection significance are evaluated. We illustrate how observation times 
vary with the cone half angle 
chosen about the Galactic center, with the result that the optimum angles are about 10$^\circ$ 
and 
50$^\circ$ for the muon events and shower events, respectively. 
We find that for 
the 
annihilating dark matter models such as the leptophilic and Kaluza-Klein models, upward and contained muon 
as well as showers are promising signals for dark matter detection in just a
few years of observation, whereas for decaying dark matter models, the same observation times
can only be reached with showers.
We also illustrate for each model 
the parameter space probed with the 2$\sigma$ signal detection in five 
years. We discuss how the shape of the parameter space probed change with 
significance and the observation time.    

\end{abstract}

\pacs{PACS: 95.35.+d, 14.60.Lm, 95.55.Vj, 95.85.Ry}

\maketitle

\section{Introduction}

The identity of dark matter (DM), which
constitutes 23$\%$ of the total density of the Universe \cite{DM_parameters},
 has been one of the most important open questions in
astrophysics for more than seven decades \cite{DM_observe}.
Concerted theoretical and
observational efforts are being
made to detect the DM through photonic and leptonic
signals
\cite{observations,PAMELA,FERMI,HESS,leptophilic,lepto,decay,
decay_Ibarra,photon_theory,gamma_simul,cirelli,KK_HESS,
gravitino_EGRET,Mft_experiments,
gravitino_PAMELA,PAMELA_FERMI_DM,kawasaki}.
Extensions of the standard model of particle physics have been proposed
to account for the observed
anomalies in the PAMELA positron
and FERMI electron plus positron
data
\cite{PAMELA,FERMI,leptophilic,decay,decay_Ibarra,
Mft_experiments,gravitino_PAMELA,PAMELA_FERMI_DM}.
These models
 also predict neutrino production via DM
annihilation or decay
\cite{neutrinos_from_DM,kumar,hisano,Covi,Covi2,boost_factor,liu,freese2}.

Neutrino signals are complementary to photon or charged lepton signals
from the Galactic center, or more generally from the Galactic halo.
Neutrinos with
energies of the order of the DM mass,
$E_\nu < m_\chi$ for $m_\chi\leq 10$ TeV,
propagate without being absorbed or deflected in transit toward the Earth.
Neutrino telescopes have the capabilities to probe higher energies than the
satellite based telescopes \cite{threshold,neutrino_telescopes}.
The approaches to DM searches, including the optimal angular coverage of the Galactic center or halo, are different depending on the details of the
DM model and the location of the detector relative to the
Galactic center. The strategies for uncovering
neutrino signals of DM decay or annihilation in the Galactic halo
seen
in  underground or underwater detectors,
are
the subject of this paper.

We calculate the neutrino induced
contained and upward muon flux, hadronic shower flux
and the muon and shower
event rates for different DM models
which are considered to explain data excesses in other indirect DM searches.
In this paper, we study the annihilation of the lightest Kaluza-Klein particle
\cite{KK_HESS,KK}, leptophilic DM particle annihilation
\cite{leptophilic,Mft_experiments},
two-body and three-body decay channels of gravitino,
as well as the decay of the leptophilic DM
\cite{decay_Ibarra,gravitino_EGRET,Mft_experiments,
gravitino_PAMELA,Covi,Covi2}.
We show that for each model, the shape of the
muon and shower fluxes
differs significantly from the shape of the
neutrino fluxes at production due to the smearing
produced by neutrino interactions and muon
propagation as discussed in Ref. \cite{erkoca,erkoca2}.
We also calculate the detection significances of the DM signals
at IceCube/DeepCore \cite{threshold} detector and compare the different DM models
in terms of the energy spectra and the
total counts of the muon and shower events.  We show the parameter space
of different DM models that is being probed when we require
2$\sigma$ muon or shower signal within five years of observation.

\section{Models for Dark Matter}

The ingredients for theoretical predictions of particle
fluxes from DM
annihilation or decay include a model for the DM
distribution, a particle
physics model for the DM particle couplings to standard model particles,
and standard model physics processes for the resulting produced particles.
For the DM distribution in the galaxy,
we use the Navarro-Frenk-White (NFW)\cite{NFW} profile as a typical realistic DM density profile.
 The expressions for the
neutrino fluxes and its dependence on the DM profile are presented in
Appendix A.
In case of DM decay (annihilation), the neutrino spectrum has linear
(quadratic) dependence on the DM density.

The particle physics
models on which we focus in this study consist of the
leptophilic, Kaluza-Klein and gravitino dark matters. Either 
thermal averaged
annihilation cross section times velocity $\langle \sigma v\rangle$ or a decay time
$\tau$ specific to the model is required.
Characteristically for annihilation,
the required  $\langle \sigma v\rangle$ is larger
\cite{Mft_experiments,kawasaki,hisano,boost_factor,liu}
than the value required for a thermal relic abundance
\cite{thermal_relic}:
$\langle \sigma v\rangle_0 = 3\times  10^{-26}\ {\rm cm}^3{\rm s}^{-1}$. Following
the current convention, we write
\begin{equation}
\langle \sigma v\rangle = B\, \langle \sigma v\rangle_0\ ,
\end{equation}
with a boost factor $B$. There are theoretical evaluations
of the boost factor \cite{boost_theory}, however, we treat the boost
factor as a phenomenological
parameter in this paper. To explain the lepton excesses,
some models have constraints on the boost factor as a
function of DM mass \cite{Mft_experiments}.

In leptophilic DM models \cite{leptophilic,Mft_experiments}
explaining the PAMELA positron excess,
the DM annihilation or decay must proceed
dominantly to leptons in order to avoid the
overproduction of antiprotons.
Moreover, according to the FERMI data, the direct production of electrons
must be suppressed with respect to the production of electrons
(and positrons) as secondaries.
It was shown \cite{Mft_experiments} that the leptophilic DM
with mass ($m_\chi$) in the range between 150 GeV and a few TeV,
which annihilates or decays into $\tau$'s or $\mu$'s
can fit the PAMELA \cite{PAMELA} and Fermi \cite{FERMI} data as well as
the HESS high energy photon data \cite{HESS}.
The best fit parameters for the boost factor
($B$) and the decay time ($\tau$)
which determine the overall normalizations, for the
specific case involving
muons from annihilation ($\chi\chi\rightarrow\mu^+\mu^-$) or
decay ($\chi\rightarrow\mu^+\mu^-$), respectively,
are given by  \cite{Mft_experiments}
\begin{eqnarray}\label{boost_mass}
B&=& 431m_\chi-38.9\nonumber\\
\tau&=&\left(2.29+\frac{1.182}{m_\chi}\right)\times10^{26} \; \mbox{sec}
\nonumber \\
&=& B_\tau \times 10^{26} \; \mbox{sec}
\end{eqnarray}
for $m_\chi$ in TeV.  The annihilation channel into tau pairs
is less favored by the data \cite{Mft_experiments}.

Some Kaluza-Klein models can provide a
DM candidate which gives the correct relic density \cite{KK}. 
To account for the   
HESS results \cite{HESS}, the
lightest Kaluza-Klein particle (LKP) would have a mass of the order
of a TeV \cite{KK_HESS}. The LKP is also assumed to be neutral and
non-baryonic.
In this model, the particle couplings are fixed such that
LKP pairs annihilate into quark pairs ($35\%$), charged lepton pairs (59$\%$),
neutrinos (4$\%$), gauge bosons (1.5$\%$) and higgs bosons (0.5$\%$)
\cite{KK_HESS,KK}.

The first DM candidate proposed in the context of supersymmetry is the gravitino ($\psi_{3/2}$)
which would be the lightest supersymmetric particle
(LSP).
The gravitino is the
superpartner of the graviton. With the existence of small R-parity breaking to
allow the LSP to decay, the gravitino decays into
standard model particles. The decay rate of the gravitino in this scenario is so small that it can have a
sufficiently long lifetime for the correct DM relic density today.

In order to   
account for the observed anomalous positron excess in the PAMELA data and
positron plus electron excess in the FERMI data, the
 lifetime of the
gravitino DM
is constrained to be of the order of
 $\sim$ 10$^{26}$ seconds and its mass to be in the range between few 100 GeV and
few TeV \cite{gravitino_PAMELA}.
To explain the data, the three-body gravitino decay mode ($\psi_{3/2}\rightarrow l^+l^-\nu$)
was considered \cite{gravitino_PAMELA}. We use the parameters of this
 model to explore neutrino signals from gravitino decay.  For illustration,
in addition to three-body decay, we also consider
the two-body gravitino decay modes ($\psi_{3/2}\rightarrow(W^\mp l^\pm ,Z\nu,\gamma\nu)$) assuming
the same lifetime and mass as for the three-body decay, and with the
 branching fractions given in Table I.

\begin{table}[h]
\begin{tabular}{cccc}
\hline
\hline
$m_{\psi_{3/2}}$(GeV) & $B_F(\psi_{3/2}\rightarrow\gamma\nu)$ & $B_F(\psi_{3/2}\rightarrow Wl)$ & $B_F(\psi_{3/2}\rightarrow Z\nu)$  \\
\hline
10 & 1 & 0 & 0 \\
85 & 0.66 & 0.34 & 0 \\
100 & 0.16 & 0.76 & 0.08 \\
150 & 0.05 & 0.71 & 0.24 \\
200 & 0.03 & 0.69 & 0.28 \\
400 & 0.03 & 0.68 & 0.29 \\
\hline
\end{tabular}
\caption{Branching fractions for the two-body gravitino decay into different R-parity violating channels
for different masses \cite{gravitino_EGRET}.}
\label{table:gravitino2}
\end{table}

\begin{table}[h]
\begin{tabular}{ccc}
\hline
\hline
Particle/mode & mass & $B_\tau$ or $B$\\
\hline
$\psi_{3/2}\rightarrow l^+ l^-\nu$ & 400 GeV & $B_\tau$=2.3\\
$\psi_{3/2}\rightarrow (Wl,Z\nu,\gamma\nu)$ & 400 GeV & $B_\tau$=2.3\\
$\chi\rightarrow \mu^+ \mu^-$ & 2 TeV & $B_\tau$=2.9\\
$B^{(1)}B^{(1)}\rightarrow (q\bar{q},l^+ l^-, W^+W^-,ZZ,\nu\bar{\nu})$ & 800 GeV & $B=200$\\
$\chi\chi\rightarrow \mu^+ \mu^-$
 & 1 TeV & $B=400$\\
\hline
\end{tabular}
\caption{Model parameters characterizing fits to explain
FERMI and PAMELA anomalies used as
examples in this paper.}   
\label{table:models}
\end{table}

\begin{figure}[h]
\begin{center}
\epsfig{file=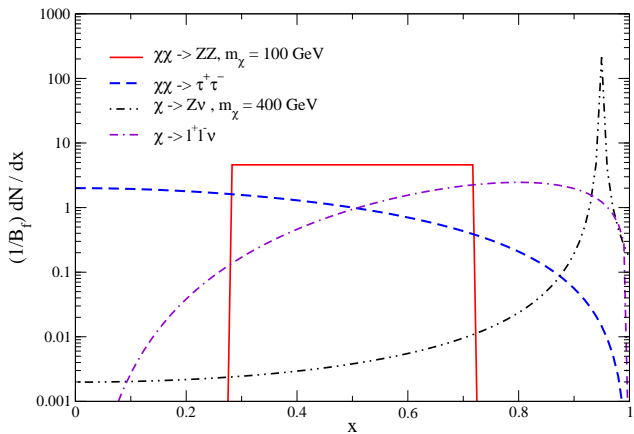,width=2.9in,angle=270}
\end{center}
\caption{Muon neutrino ($\nu_\mu$) spectra in terms of $x=E_\nu/E_{\nu,max}$
from
the three-body decay of gravitino (dot-dash-dashed line), from the decay of
$\tau$ (dashed line), $Z$ boson (solid line) and
from one of the two-body decay channels of gravitino
($\psi\rightarrow Z\nu$) for which Breit-Wigner distribution is used.
The distributions
should
be multiplied by the branching fractions, and oscillations should be taken into account for the flux of neutrinos at Earth.}
\label{fig:enudist}
\end{figure}

Selected DM model parameters are shown in Table \ref{table:models}.
For each of the DM models considered, the decay distribution of the produced
particles to neutrinos in case of DM annihilation, or the gravitino decay distribution
to neutrinos, enters into the calculation of the neutrino fluxes that arrive at Earth.
For annihilation directly to neutrinos, the energy distribution of each neutrino is
a delta function in energy, with the energy equal to the DM mass. This case has
been well studied in the literature 
\cite{kumar,erkoca,erkoca2,direct_neutrinos}.
Here, we look at the secondary neutrinos.
Fig. \ref{fig:enudist} shows neutrino spectra,
 plotted in terms of $x\equiv E_\nu/E_{\nu,max}$ where $E_{\nu,max}=m_\chi$
for annihilating DM and $E_{\nu,max}=m_\chi/2$ for decaying DM
models. The curves in the figure are
normalized to count the number of neutrinos, and in the case of the   
$Z\nu$ final state, the fraction of $Z$ decays to neutrinos. The
muon neutrino spectra in the figure should be multiplied
by the branching fraction for a specific decay channel in a given model.
Analytic expressions
for the neutrino spectra are given in Appendix B.

\section{Neutrino Flux}

\begin{figure}[h]
\begin{center}
\epsfig{file=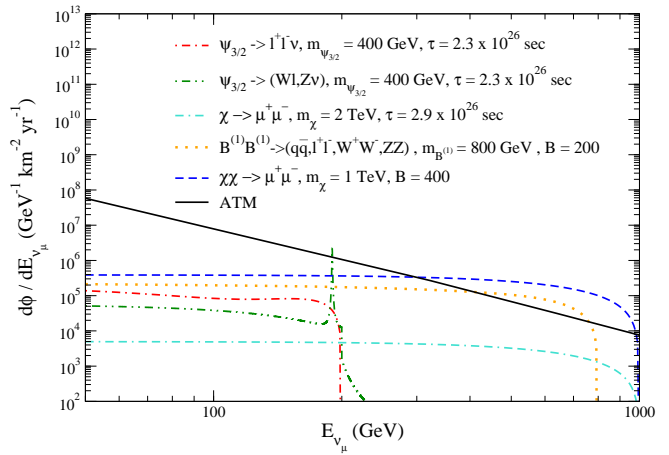,width=2.9in,angle=270}
\end{center}
\caption{Muon neutrino ($\nu_\mu$) fluxes from the annihilation of the Kaluza-Klein (dotted line), leptophilic (dashed line),
 and the decay of leptophilic (dash-dot-dashed line), three-body decay (dot-dashed line)
and two-body decay (dot-dot-dashed line) of gravitino DM particles. Neutrino oscillations have been
taken into account.
The angle-averaged atmospheric muon neutrino flux
at the surface of the Earth is also presented (solid line). The corresponding values of the parameters
for each model are shown as well.}
\label{fig:munuspectra}
\end{figure}

The neutrino flux at Earth can be evaluated using the neutrino flux expressions 
given by Eqs. (\ref{annihilation_flux}) and (\ref{decay_flux}) 
in Appendix A and the neutrino spectra given in Appendix B with 
taking the 
neutrino oscillation effects into account.
In Fig. \ref{fig:munuspectra}, we show the muon neutrino flux at Earth
for three decay channels and two annihilation channels. 

In some DM models all three flavors of neutrinos can be generated by DM annihilation or decay,
implying the flavor ratio at the production to be $\nu_e:\nu_\mu:\nu_\tau$ is 1:1:1.  This ratio remains
unchanged with oscillation.  This is the case for the gravitino decay and Kaluza-Klein
DM annihilation. However, in case of the leptophilic DM model,
in which
$\chi \rightarrow \mu^+\mu^-$,
the initial neutrino flavor
ratio is $1:1:0$
which becomes $1:0.5:0.5$ as neutrinos travel astrophysical distances. We take this
oscillation effect into
account when we evaluate muon neutrino fluxes presented in
 Fig. \ref{fig:munuspectra} and when we evaluate muon event rates below.

Fig. \ref{fig:munuspectra} shows that with the exception of the gravitino decays, the distributions of
neutrinos have very weak energy dependence.
 The two-body gravitino decay gives a spiked feature at the kinematic limit
in neutrino energy. The relative normalizations of the DM curves comes from different DM lifetimes or boost factors.

Also shown in Fig. \ref{fig:munuspectra} is the angle-averaged atmospheric 
muon neutrino
flux at the surface of the Earth. It is characterized by an approximate formula
\cite{erkoca,neutrino_background1},
(in units of GeV$^{-1}$km$^{-2}$yr$^{-1}$sr$^{-1}$)
\begin{eqnarray}\label{background}
\left(\frac{d\phi_\nu}{dE_\nu d\Omega}\right)_{ATM,avg} &=& N_0{E_\nu}^{-\gamma-1}\Biggl(\frac{a}{bE_\nu}\ln(1+bE_\nu)+\nonumber\\
 &+ &\frac{c}{eE_\nu}\ln(1+eE_\nu)\Biggr)\ ,
\end{eqnarray}
where the parameters in the formula are listed in Table \ref{table:atm}.
These same parameters appear in the
angle-dependent atmospheric neutrino flux for zenith angle $\theta$,
\begin{eqnarray}\label{background2}
\frac{d\phi_\nu}{dE_\nu d\Omega} &=&
N_0{E_\nu}^{-\gamma-1}\nonumber\\
&\times &\left(\frac{a}{1+bE_\nu{cos\theta}}+\frac{c}{1+eE_\nu{cos\theta}}\right)\ .
\end{eqnarray}
This formula does not account for the prompt neutrino flux \cite{prompt_neutrino}, however, for the energy range of interest, the prompt
atmospheric
neutrino flux is negligible.

\begin{table}[t]
\begin{tabular}{|c|c|} 
\hline
\hline
$\gamma $& 1.74\\
$a$ & 0.018\\
$b$ & 0.024\ GeV$^{-1}$\\
$c$ & 0.0069\\
$e$ & 0.00139 GeV$^{-1}$\\
$ N_0$ & $
         \begin{array}{lr}
         1.95\times10^{17}&  \mbox{for}\;\;\nu \\
         1.35\times10^{17}&  \;\mbox{for}\;\;\overline\nu.
         \end{array}$
         \\
         \hline
         \hline 
         \end{tabular}
\caption{Parameters for the atmospheric $\nu_\mu$ and $\bar{\nu}_\mu$ fluxes given by Eqs. (\ref{background}) and (\ref{background2}),
in units of
GeV$^{-1}{\rm km}^{-2}{\rm yr}^{-1}{\rm sr}^{-1}$ \cite{neutrino_background1}.}
\label{table:atm}
\end{table}

\begin{figure}[h]
\begin{center}
\epsfig{file=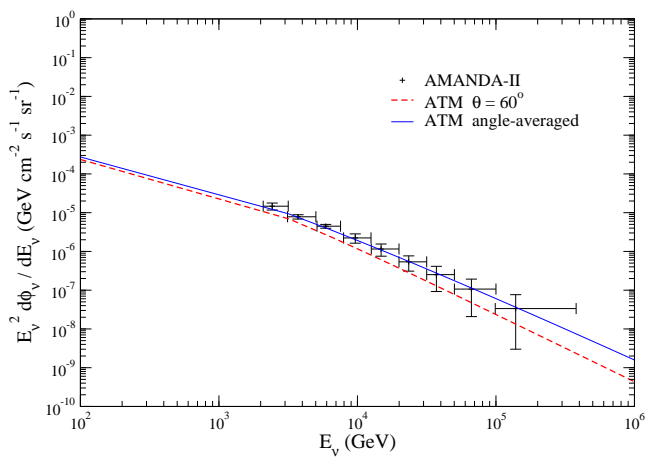,width=2.9in,angle=270}
\end{center}
\caption{Angle-averaged atmospheric muon neutrino ($\nu_\mu+\overline{\nu}_\mu$) flux (solid line) 
and the atmospheric flux for fixed
$\theta=60^{\circ}$ (dashed
line) compared with the angle-averaged ($\nu_\mu+\overline{\nu}_\mu$) flux from
AMANDA-II muon neutrino flux measurements \cite{AMANDA_data}.
}
\label{fig:amandadata}
\end{figure}

The angle-averaged atmospheric neutrino flux is a good approximation. In Fig. \ref{fig:amandadata},
we show the angle-averaged flux from Eq. (4) and the flux from Eq. (5) with $\theta=60^\circ$ and the
integrated flux measured
by the AMANDA-II detector from Ref. \cite{AMANDA_data}. The angle-averaged flux is a bit larger
than the flux at $60^{\circ}$, so at least for  $\theta$ less than $60^\circ$, using the angle-averaged
atmospheric flux gives a small overestimate of the atmospheric background.

\section{Neutrino signals}

For each DM candidate and model, there are several
signals to pursue in underground detectors. One possibility is to measure or constrain
the rate of muons produced by muon neutrinos, over and above the expected atmospheric background rate.
High energy muons point essentially in the same direction as the incident neutrino,
and the angular resolution of high energy muon tracks is quite good. With good enough
energy and angular resolution, and a large enough target volume, one looks for neutrinos
coming directly from DM annihilation in the Galactic center, however,
the target volume may be a limitation for constraining
model parameters including the boost factor.
A comparison of the upward-going muon rate, where the target volume is enhanced
by the muon range at high energies, and the contained rate of muon production by neutrinos in the
detector, is a useful exercise.

For the IceCube/DeepCore detector, the Galactic center is above the horizon, so
the upward muon rate of DM produced neutrinos is from the Galactic halo in a direction pointing away from the Galactic center. We consider this
possibility as
well.

Showers, either electromagnetic or hadronic, are produced by neutrinos. 
We look at the optimization for these as well as a function of cone half angle,
but we note that the current capabilities for shower angular resolution are
somewhat limited, on the order of 50$^\circ$ \cite{volume_eff,middell}.

\subsection{Muons}

The neutrinos coming from the
Galactic center and Galactic halo can produce muons through charged current interactions
in the detector (contained muons). The flux is given by
\begin{eqnarray}\label{galactic_flux}
\frac{d\phi_\mu}{dE_\mu}&=& \int^{E_{max}}_{E_\mu}dE_\nu
\left(\frac{d\phi_\nu}{dE_\nu}\right)\frac{N_A\rho}{2}\times\nonumber\\
&\times&\left(\frac{d\sigma^p_\nu(E_\nu,E_\mu)}{dE_\mu}+(p\rightarrow{n})\right) + (\nu\rightarrow \bar{\nu}).
\end{eqnarray}
where $N_A =6.022\times10^{23}$ is Avogadro's
number, $\rho$ is the density of the medium, $E_{max}=m_\chi$ for annihilation and
$E_{max}=m_\chi/2$ for decay.  The differential cross
sections
$d\sigma_{\nu}^{p,n}/dE_\mu$ are the weak scattering charged-current cross sections for
neutrino and antineutrino scattering with protons
and neutrons \cite{weak_scatter}.

\begin{figure}[h]
\begin{center}
\epsfig{file=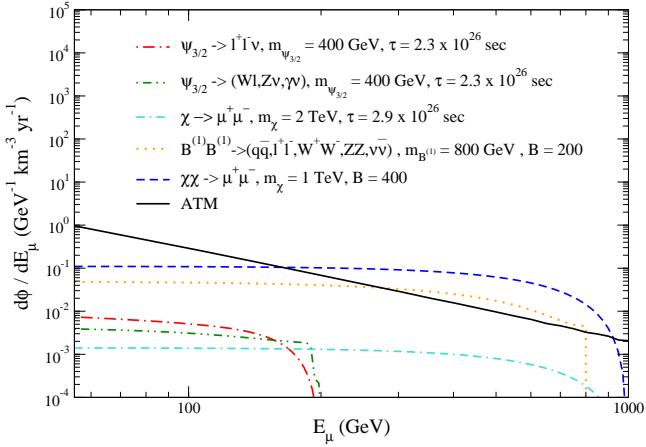,width=2.9in,angle=270}
\end{center}
\caption{Muon flux for the contained events for gravitino decay
(dot-dashed and dot-dot-dashed lines), Kaluza-Klein annihilation
(dotted line), leptophilic model (dashed line for annihilation and
dash-dash-dotted line for decay) compared with the atmospheric
background (solid line), for the case when
 $\theta_{{\rm max}}=1^\circ$. Model parameters are given in Table II.}
\label{fig:containedmu}
\end{figure}

We evaluate the muon
flux from neutrino charged-current interactions in the detector,
when neutrinos are produced in
DM annihilation or DM decay.  In
Fig. \ref{fig:containedmu} we show muon fluxes for the
case when the DM particle is a gravitino, a Kaluza-Klein
particle and for a leptophilic model in which
 DM annihilation or decay produces  $\mu^+\mu^-$, for the model parameters listed in Table II.
We take the cone half angle around the Galactic center
to be
$\theta_{\rm max} = 1^\circ$.

In case of the gravitino DM decay and for Kaluza-Klein DM
annihilation, there are
 discontinuities in the slopes at the highest muon energies
coming from the superposition of the
direct neutrino production (dot-dot-dashed line for DM decay and
dotted line for
the Kaluza-Klein annihilation).
 The direct neutrino production,
 $\chi\chi\rightarrow\nu\bar{\nu}$ is the ``golden channel'' for DM detection
because in this case the muon flux is increasing with energy, and it peaks at
$E_\mu = m_\chi$ \cite{erkoca2}.

As noted in Sec. II,
the parameters used for DM masses, boost factors and lifetimes are
characteristic of those that
were shown to describe PAMELA, Fermi/LAT and HESS data
\cite{KK_HESS,Mft_experiments,gravitino_PAMELA}.
Changing the value of the boost factor or the lifetime affects only the
overall normalization of the muon flux.
We find that for this choice of
the parameters,
DM signals in leptophilic model exceed the atmospheric
background for $E_\mu > 175$ GeV, while for the
Kaluza-Klein DM model the signal
is above the background for
$E_\mu> 275$ GeV.  In both cases, the signal cuts off when
 $E_\mu = m_\chi$.
 
\begin{figure}[h]
\begin{center}   
\epsfig{file=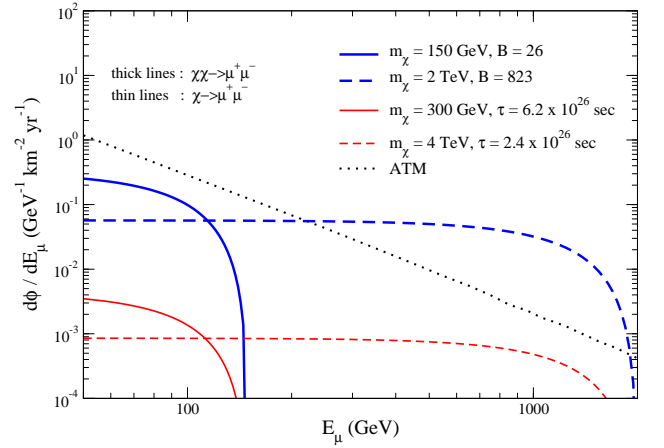,width=2.9in,angle=270}
\end{center}
\caption{Contained muon flux from the annihilation, $\chi\chi\rightarrow\mu^+\mu^-$
(thick lines) and the decay, $\chi\rightarrow\mu^+\mu^-$ (thin lines) processes. The
relation between the boost factor and $m_\chi$, and   
between the lifetime and $m_\chi$ are given by Eq. (\ref{boost_mass}).
We take $\theta_{\rm max} = 1^\circ$.
}
\label{fig:containedlepto}
\end{figure}   

We consider the effect on the muon flux shape when we change the
 parameters, for example for the
leptophilic model.  In
 Fig. \ref{fig:containedlepto} we show the contained muon 
flux from DM annihilation and
decay in a leptophilic model for different values of
the parameters $B$, $\tau$ and $m_\chi$, which are constrained to
satisfy Eq. (\ref{boost_mass}) to describe
the data \cite{Mft_experiments}.
The decays (lower thin lines) have lower fluxes than the annihilations (upper thick lines), even though
the shapes are similar. For leptophilic models, one cannot enhance the signal rate by
 increasing $B$ or decreasing $\tau$ with $m_\chi$ fixed
 if the Fermi and PAMELA data are explained by the model.

 Contained muons, produced by neutrino interactions in the
detector, make up one set of muon signals.
Muons can also be produced in
neutrino interaction in
the rock below the detector. Muons produced
 with energy $E^i_\mu$, interact with the medium
and finally reach the detector  with energy $E_\mu$.  The
effective volume of the detector is enhanced by the
muon range at high energies.  We denote these events
as upward muon events.

The muon range in the rock,
$R_\mu(E_\mu^i,E_{\mu})$, depends on the initial muon energy $E_\mu^i$,
the final energy $E_{\mu}$ and the parameters $\alpha$ and $\beta$ which
characterize muon energy loss. Numerically,
$\alpha \simeq 2\times 10^{-3}$ GeV\,cm$^2$/g accounts for
the ionization energy
loss and $\beta \simeq 3.0\times 10^{-6}$ cm$^2$/g for the
bremsstrahlung, pair production and photonuclear interactions. The range is then
approximated by $R_\mu(E_\mu^i,E_\mu) = \ln\bigl[(E_\mu^i+\alpha/\beta)/
(E_\mu+\alpha/\beta)\bigr]/\beta\rho$.  For muon transit through the 
rock,
the muon range is 1 km for $E_\mu^i\simeq 1$ TeV.

Taking into account the energy losses, the final muon flux at the position of the detector can be written as \cite{erkoca},
\begin{eqnarray}\label{galactic_flux2}
\frac{d\phi_\mu}{dE_\mu}&=&\int^{R_\mu(E_\mu^i,E_\mu)}_0 e^{\beta\rho 
z}dz \int^{E_{max}}_{E^i_\mu}dE_\nu
\left(\frac{d\phi_\nu}{dE_\nu}\right)\nonumber\\
&\times & 
P_{surv}(E^i_\mu,E_\mu)\frac{N_A\rho}{2}\left(\frac{d\sigma^p_\nu(E_\nu,E_\mu)}{dE_\mu}+(p\rightarrow{n})\right) 
+\nonumber\\ &+& (\nu\rightarrow \bar{\nu}) ,
\end{eqnarray}
where $P_{surv}(E^i_\mu,E_\mu)$
is the survival probability for a muon with initial energy
$E^i_\mu$ to reach final energy $E_\mu$.   
A detailed derivation of Eq. (6) can be found in Ref. \cite{erkoca}.
To first approximation,
$P_{surv}\simeq 1$
since at high
energies, the muon has a long decay length. The muon energy at the production point is related to the muon energy a distance
$z$ from that point by
\begin{equation}
E^i_\mu(z) \simeq e^{\beta \rho z}E_\mu +(e^{\beta\rho 
z}-1)\frac{\alpha}{\beta}\ .
\end{equation}

\begin{figure}[h]
\begin{center}
\epsfig{file=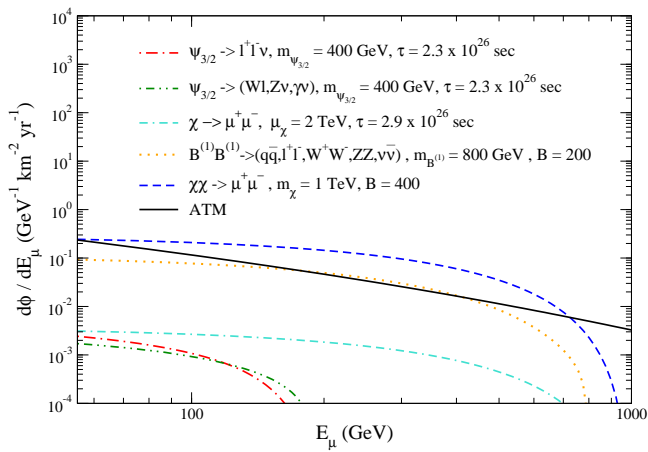,width=2.9in,angle=270}
\end{center}
\caption{Upward muon flux for the annihilating and decaying DM models from
Table II. We take $\theta_{\rm max}=1^\circ$.}
\label{fig:upmu}
\end{figure}

We show in Fig. \ref{fig:upmu}, the upward muon flux for a
generic northern hemisphere detector,
looking down through the Earth with a cone half angle of
$\theta_{\rm max}=1^\circ$ around the Galactic
center. The muon fluxes are smoothed relative to Fig. \ref{fig:containedmu} as
a consequence of the energy loss. For DM particles with masses
of order 1 TeV, the upward muon flux is larger at low energies than for the contained
muons because the muon range is larger than 1 km, effectively enhancing the
volume of the kilometer-size detector. When $m_\chi = 400$ GeV, the energies of the produced muons are
such that the muon range is less than 1 km, which is the size of the detector for which
the contained muon flux was calculated. If the depth of the detector is 500 m,
the contained and upward muon fluxes for the DM mass of 400 GeV would be
approximately equal at low energies, although the contained muon flux would have
a little harder spectrum.  This is direct consequence of muon range dependence
on the DM mass.  For example, muon with initial energy of 400 GeV (1 TeV) 
has a range of 500m (1 km).

The contained muon event rate, $N^{ct}_{\mu}(m_\chi)$,
 is obtained by integrating muon flux folded
with the effective volume of the detector,
$V_{\rm eff}$, i.e.
\begin{equation}\label{contained_event}
N^{ct}_\mu(m_\chi)=\int^{E_{max}}_{E^{th}_\mu} \frac{d\phi^{ct}_\mu}{dE_\mu} V_{\rm eff} (E_\mu)
dE_\mu
\end{equation}
where ${d\phi^{ct}_\mu}/{dE_\mu}$ is given in
Eq. (\ref{galactic_flux}) and
$E^{th}_\mu$ is the muon detector threshold, typically 10-100 GeV for deep ice or
water detectors \cite{threshold}. In our calculations, we choose $E^{th}_\mu=50$ GeV.
We also consider an energy independent IceCube/DeepCore effective volume,
$V_{\rm eff}=0.04$ km$^3$, for the contained muon events
\cite{freese2,volume_eff}.

Similarly,
the upward muon event rate,
is obtained by
\begin{equation}\label{upward_event}
N^{up}_\mu(m_\chi)=\int^{E_{max}}_{E^{th}_\mu} \frac{d\phi^{up}_\mu}{dE_\mu} A_{\rm eff} (E_\mu)
dE_\mu
\end{equation}
where  ${d\phi^{up}_\mu}/{dE_\mu}$ is given by
Eq. (\ref{galactic_flux2}),
$ A_{\rm eff}$ is the angle-averaged muon
effective area for which we assume
$A_{\rm eff}=1$ km$^2$.  

The event rates for contained
and upward muons for a cone half angle of 
$\theta_{\rm max}=1^\circ$ 
are shown in Table
\ref{table:muonevtrates} for the DM models shown in Figs.
\ref{fig:containedmu} and \ref{fig:upmu} and listed in Table II.
We also obtain the number of years
required for the rate of signal events $s$ and background events $b$ to satisfy the condition
\begin{equation}
\label{eq:signal}
\frac{s}{\sqrt{s+b}}\geq 2\ .
\end{equation}
From Table \ref{table:muonevtrates} we note that for the parameters that
we considered,
 only the Kaluza-Klein
DM and leptophilic annihilation models have a reasonable chance
of detection for $\theta_{\rm max}=1^\circ$.

\begin{table}[h]
\centering
\begin{tabular}{l|cc|cc}
\hline\hline
  &\multicolumn{2}{c}{$ A_{\rm eff} =1$km$^2$} & \multicolumn{2}{c}{$ V_{\rm eff} =0.04$ km$^3$} \\
\hline
& N$^{up}_\mu$ & t (yr) & N$^{ct}_\mu$ & t (yr) \\
\hline
$\psi_{3/2}\rightarrow$l$^+$l$^-\nu$ & 0.12 & 7811  & 0.0224 & 1.8$\times10^4$ \\
$\psi_{3/2}\rightarrow (Wl,Z\nu$,$\gamma\nu$) & 0.1 & 1.1$\times10^4$ & 0.0156 & 3.8$\times10^4$ \\
$\chi\rightarrow\mu^+\mu^-$ & 0.6 & 317 & 0.027 & 1.2$\times10^4$ \\
$B^{(1)}B^{(1)}\rightarrow (q\bar{q},l^+l^-,\nu\bar{\nu},...)$ & 16 & 0.7 & 0.72 & 23 \\
$\chi\chi\rightarrow\mu^+\mu^-$ & 46 & 0.14 & 2.1 & 4 \\
\hline
ATM & 28 &  & 2.28 & \\
\hline
\end{tabular}
\caption{
Event rates per year and the time required to reach 2$\sigma$ detection significance for the
upward and the contained muons
($\mu$) for a cone half angle of $\theta_{\rm max}=1^\circ$. Results for different DM models
 are obtained by taking $A_{\rm eff}=1$km$^2$ and
$V_{\rm eff}=0.04$ km$^3$ for the upward and contained muon events, respectively.
}
\label{table:muonevtrates}
\end{table}

The model parameters such as DM masses, annihilation cross sections and
decay times that we consider are introduced to
explain some indirect DM searches as explained in the previous section. However,
it is also possible that the signals that have been observed 
\cite{PAMELA,FERMI,HESS} in these searches
have no DM origin.
Then, $m_\chi$ and $B$ or $\tau$ can be varied independently.
In terms of the neutrino signals,
the dependence of the signals on the annihilation cross sections or on the decay times is
trivial since these parameters affect only the overall normalization.
The dependence on DM mass
is not that straightforward.

\begin{figure}[h]
\begin{center}
\epsfig{file=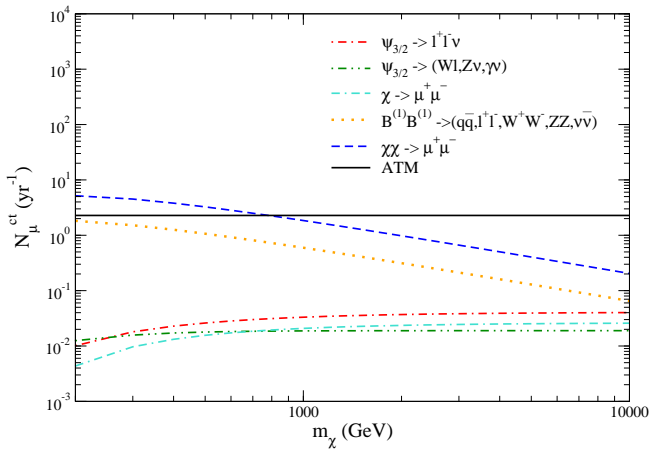,width=2.9in,angle=270}
\end{center}
\caption{Contained muon rates as a function of DM mass for the models presented in Table II.
We take the muon detector threshold to be $E^{th}_\mu=50$ GeV
and the cone half angle to be $\theta_{\rm max}=1^\circ$.}
\label{fig:containedmu_rates}
\end{figure}

\begin{figure}[h]
\begin{center}
\epsfig{file=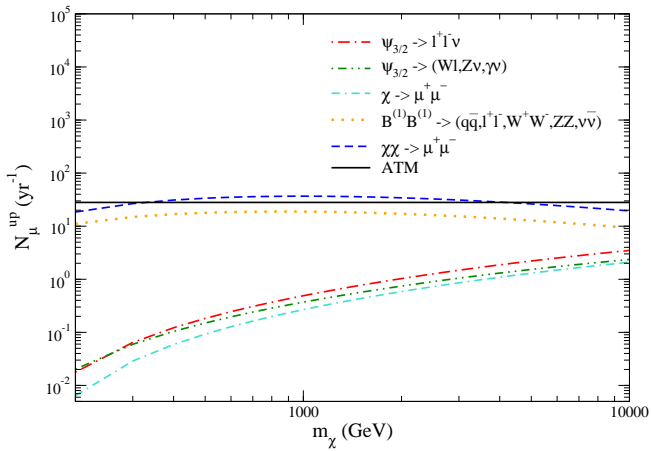,width=2.9in,angle=270}
\end{center}
\caption{ Upward muon rates as a function of DM mass for the DM models presented in Table II.
We take the muon detector threshold to be 
$E^{th}_\mu=50$ GeV
and the cone half angle to be $\theta_{\rm max}=1^\circ$.}
\label{fig:upmu_rates}
\end{figure}

In order to see the dependence of the signals
 on DM mass, we set the values of the
boost factor and the decay times to those in
Table II and calculate the event rates as a function of   
DM mass for each model. We present our results for the contained
muon event rates in Fig.
 \ref{fig:containedmu_rates}
and for the
upward muons in Fig. \ref{fig:upmu_rates}.
The solid line in each figure corresponds to the muon background
due to the atmospheric neutrinos.

From Fig. \ref{fig:containedmu_rates} we note that the contained
muon events rates
for
annihilating DM models
 decrease with $m_\chi$.  On the other hand
the event rates
for the decaying DM models
increase slowly with $m_\chi$, for $m_\chi<1$ TeV
and for $m_\chi>1$ TeV, they become almost independent of $m_\chi$.
The $m_\chi$ dependence of the contained muon event rates
is mainly due to the
$m^{-2}_\chi$ ($m^{-1}_\chi$) dependence in the
neutrino flux
for DM annihilation (DM decay) combined with the upper limit of
integration dependence on $m_\chi$.
For DM masses in the range
$E^{th}_\mu < m_\chi < $ 400 GeV, where
$E^{th}_\mu = 50$ GeV, the integration region is sensitive to the
value of the DM mass, while
for $m_\chi\gg E^{th}_\mu$ there is only weak dependence on $m_\chi$.
These combined effects are responsible for the observed
$m_\chi$ dependence of the contained muon event rates presented
in Fig. \ref{fig:containedmu_rates}.
 
The $m_\chi$ dependence of the upward muon rates is shown in
Fig. \ref{fig:upmu_rates}.  We find that
 the event rates for decaying DM models increase with $m_\chi$ while
for annihilating DM models there is almost no $m_\chi$ dependence for
a wide range of DM masses.
In contrast to the contained muon rates,
for upward muons
there is additional $m_\chi$ that is present
 in the muon range.
As we increase the value of DM mass, the effective volume
which depends on the muon range in rock becomes larger.

The upward muon rates for a decaying DM particle have
 steeper increase with increasing DM mass than for
contained muon rates, because of
the energy dependent
effective volume which increases with $m_\chi$,
 when compared to the case for the contained muon events.
 
\begin{figure}[h]
\begin{center}
\epsfig{file=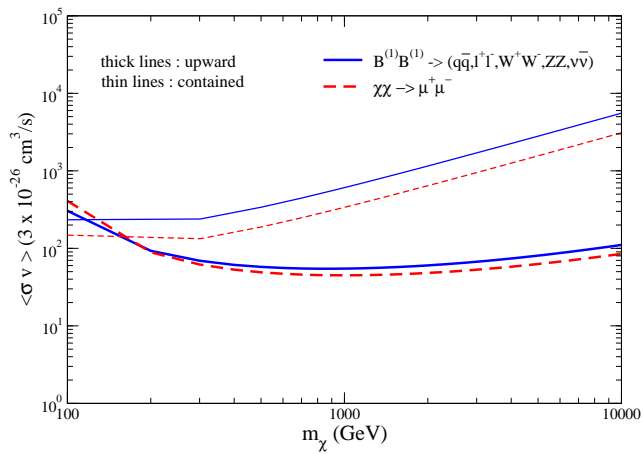,width=2.9in,angle=270}
\end{center}
\caption{Annihilation cross section versus DM mass for upward and for
contained muon events to reach 2$\sigma$ detection significance with  
 five years of observation for the
case of Kaluza-Klein (solid lines) and leptophilic (dashed lines) models.
We take $\theta_{\rm max}=1^\circ$ and $E^{th}_\mu=50$ GeV.
}
\label{fig:boost_mass1}
\end{figure}

In Fig. \ref{fig:boost_mass1}, we present results for
DM annihilation cross section required for a given
DM mass
in order to reach 2$\sigma$ detection significance in five years of observation
within $\theta_{\rm max}=1^\circ$ for Kaluza-Klein (solid lines) and annihilating leptophilic (dashed lines) models.
From Fig. \ref{fig:containedmu_rates} we note that
the contained muon event rates decrease with $m_\chi$ for a
fixed annihilation cross section (i.e. fixed boost factor).
Therefore,
in order to have the same detection significance for each DM mass,
DM annihilation cross section needs to increase with $m_\chi$ as shown
in Fig. \ref{fig:boost_mass1}.
However, for the upward muons, the event rates increase with $m_\chi$ for $m_\chi<1$ TeV and exhibit a slight decrease for higher
DM masses for a fixed annihilation cross section. Thus, in order to have the same significance independent of the
DM mass for the upward muon events the DM annihilation cross section has to decrease with $m_\chi$ for $m_\chi<1$ TeV and increase for
$m_\chi>1$ TeV as seen in Fig. \ref{fig:boost_mass1}.
If there is no signal detected at 2$\sigma$ level in five years, 
  the parameter space above each curve is excluded 
at that significance level. Our results also indicate that
the upward muons are more promising than the contained muons
in constraining the model parameters.
Increasing the observation time would result in larger
excluded parameter space.

\begin{figure}[h]
\begin{center}
\epsfig{file=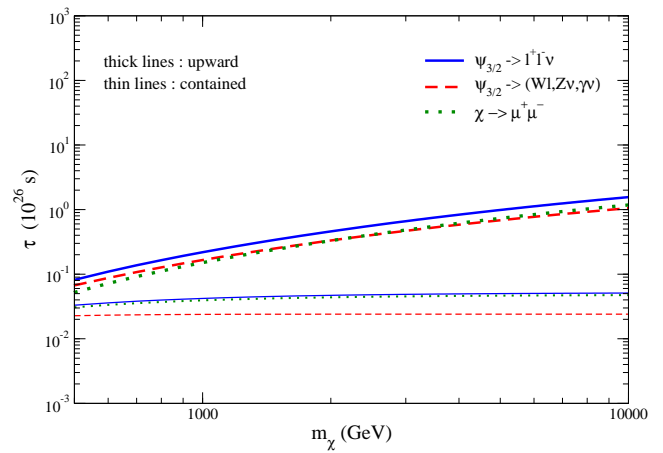,width=2.9in,angle=270}
\end{center}
\caption{Decay time versus DM mass for upward and for
contained muon events to reach 2$\sigma$ detection significance with
 five years of observation for the
case of decaying DM models: gravitino three-body(solid lines), gravitino two-body (dashed lines)
and leptophilic (dotted lines).
We take $\theta_{\rm max}=1^\circ$ and $E^{th}_\mu=50$ GeV.
}
\label{fig:decayt_mass1}
\end{figure}

Similar to the models for the
annihilating DM, we evaluate the parameter space for the decaying DM models.
In Fig. \ref{fig:decayt_mass1} we show the
decay time as a function of the DM mass for $\theta_{\rm max}=1^\circ$ that is needed
in order to reach 2$\sigma$ detection significance
with five year observation period.
For a wide range of DM masses, the contained muon event rates
have weak dependence on $m_\chi$, while
 the upward muon event rates
show a steep increase with increasing $m_\chi$ as can be seen from Figs. \ref{fig:containedmu_rates}
and \ref{fig:upmu_rates}.
This implies that we need longer
decay time for the upward muon events than for the contained muon events to reach the same detection
significance for a five year observation time while
 the decay time has almost no dependence on the DM mass for the contained muon events.
The parameter space below each curve corresponds to the
exclusion region at 2$\sigma$ level after five years of no signal detection.

\subsection{Optimal angles and muon signals}

\begin{figure}[h]
\begin{center}
\epsfig{file=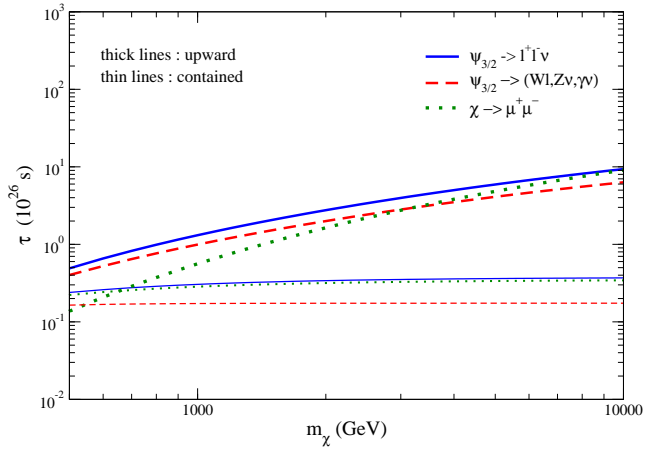,width=2.9in,angle=270}
\end{center}
\caption{Decay time versus DM mass for both upward and
contained muon events to reach 2$\sigma$ detection significance after five 
years of observation for the
case of decaying DM models: gravitino three-body (solid lines), gravitino two-body (dashed lines)
and leptophilic (dotted lines).
We take $\theta_{\rm max}$ = 10$^\circ$ and $E^{th}_\mu=50$ GeV.
}
\label{fig:decayt_mass2}
\end{figure}

As noted earlier, size of the cone (cone half angle value) 
that is too small can restrict the number of events too much.
In Fig. \ref{fig:decayt_mass2},
we present contour plots similar to those in
Fig. \ref{fig:decayt_mass1} but for $\theta_{\rm max}=10^\circ$.
With increasing $\theta _{max}$ from 1$^\circ$ to
10$^\circ$,
the $J$ factor ($\langle J_1\rangle_\Omega \Delta \Omega$)
entering into the neutrino flux increases,
resulting in stronger restriction on the DM lifetime.

For more general use in discussing the dependence on cone half 
angles, the 
values of
$\langle J_1\rangle_\Omega \Delta \Omega$ and
$\langle J_2\rangle_\Omega \Delta \Omega$ given in Table VI
in Appendix A  can be combined with the
parametric forms for the muon fluxes given in Appendix C  to calculate
event rates for different cone half angles around the Galactic center.
From Table VI we note that the cone size dependence of the $J$ factors
is quite different for the case of DM annihilation than for
decaying DM.

The difference between DM density contributions to DM annihilation and decay
can also be seen in Fig. \ref{fig:J_factors} in Appendix A, where we present $\langle J_n\rangle_\Omega\Delta\Omega$ factors for
DM annihilation
($n=2$) and DM decay ($n=1$) evaluated
for a cone wedge between
$\theta_{\rm max}-1^\circ$ and $\theta_{\rm max}$ around the Galactic
center. We note that moving the wedge away from
the galactic center
leads to a significant reduction of
the signal from annihilating DM due to the dependence of
$\langle J_2\rangle_\Omega \Delta \Omega$ on
square of the density (see Eq. (\ref{J_factor_n})). We find that the
signal is reduced by a factor of 17
when the wedge is
moved from 1$^\circ$ to 90$^\circ$ off the galactic center.  On the other
hand,
in a wedge between $50^\circ$ and $90^\circ$ around
the Galactic center, the value of
$\langle J_2\rangle_\Omega \Delta \Omega$
is about 5.6, a value close to what one obtains for
a cone centered at the Galactic center with $\theta_{\rm max} = 5^\circ$.

In contrast, for a decaying DM, the signal from a wedge between
$\theta_{\rm max}-1^\circ$ and $\theta_{\rm max}$ around the Galactic
center increases with $\theta_{\rm max}$ for
$\theta_{\rm max}<30^\circ$ and decreases slowly for higher
$\theta_{\rm max}$.  For example,
$\langle J_1(1^\circ)\rangle_\Omega \Delta \Omega=0.018$ whereas
$\langle J_1(30^\circ)-J_1(29^\circ)\rangle_\Omega \Delta \Omega=0.2$
which is only a factor of 2 higher than
  $\langle J_1(90^\circ)-J_1(89^\circ)\rangle_\Omega \Delta \Omega=0.1$.
This is a consequence of the dependence of the signal for a decaying
DM particle on   
density which is one power less when compared to an annihilating DM
particle. An increase in the
size of the wedge (i.e., the volume of the source region) can result in an
enhancement in the
signal even if the chosen wedge is away from the Galactic center.

\begin{figure}[h]
\begin{center}
\epsfig{file=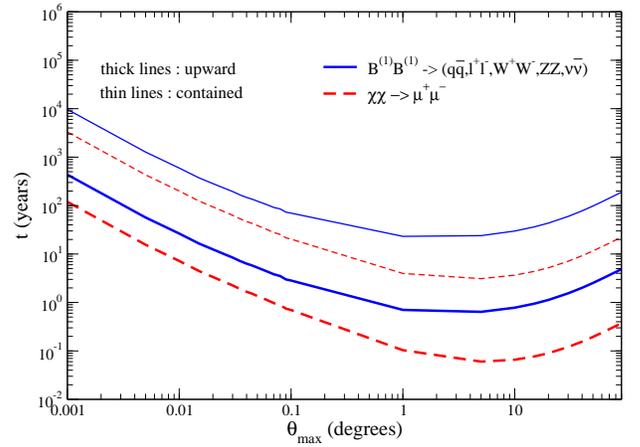,width=2.9in,angle=270}
\end{center}
\caption{The observation time $t$ versus cone half angle $\theta_{\rm max}$ centered at the
Galactic center to reach
2$\sigma$ detection significance for the upward and contained muon events produced
by the muon neutrinos originated from the annihilation of DM particles. Kaluza-Klein (solid lines) and
leptophilic where
$\chi\chi\rightarrow\mu^+\mu^-$ (dashed lines) models are considered. We take $E^{th}_\mu=50$ GeV.
}
\label{fig:up_sigma}
\end{figure}     

\begin{figure}[h]
\begin{center}
\epsfig{file=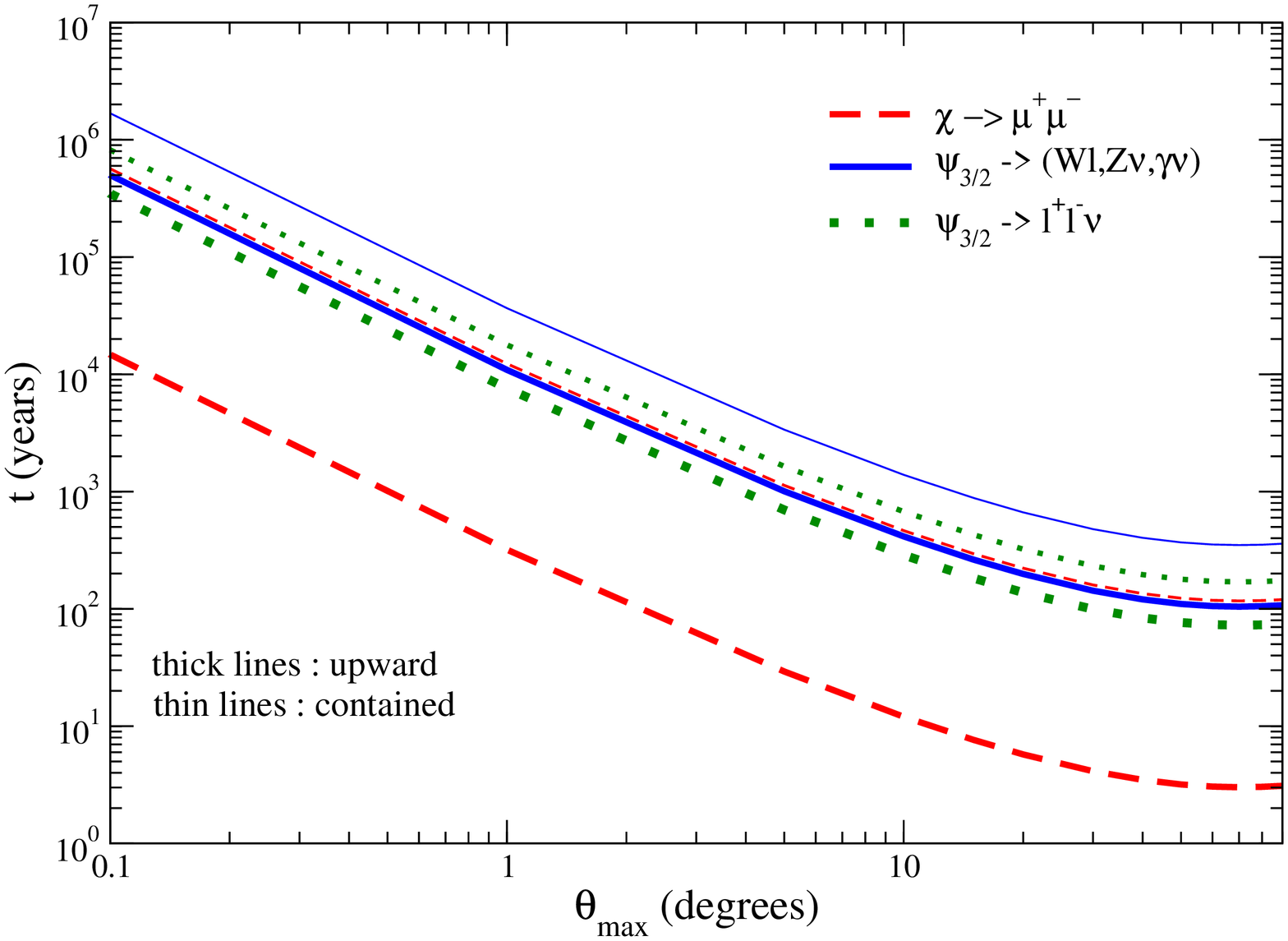,width=2.9in,angle=270}
\end{center}
\caption{The observation time $t$ versus cone half angle $\theta_{\rm max}$ centered at the
Galactic center to reach
2$\sigma$ detection significance for the upward and contained muon events produced
by the muon neutrinos originated from the decaying DM particles. Decay of leptophilic where
$\chi\rightarrow\mu^+\mu^-$ (dashed lines), two-body decay (solid lines) and three-body decay (dotted lines)
of gravitino are considered. We take $E^{th}_\mu=50$ GeV.}
\label{fig:up_sigma2}
\end{figure}

In what follows we consider cone half angles larger than $\theta_{\rm max}=1^\circ$.
In Fig. \ref{fig:up_sigma},
the detection times required for the signal $s$ to satisfy Eq. (\ref{eq:signal}) for the Kaluza-Klein
and leptophilic annihilation
cases are shown as a function of cone half angle. The optimal angle 
depends on the model and is
of the order of a few degrees.
The observation time is almost independent on the cone half angle 
in the range of a few degrees
to about 10 degrees, but it increases
for larger cone half angles where the signal increases slower than
the background.

As shown in Fig. \ref{fig:up_sigma2} for the decaying DM models,
the observation time to reach the same significance
decreases with the cone half angle
 $\theta_{\rm max}$, for
$\theta_{\rm max} <50^\circ$, and it increases only slowly for the higher cone sizes, as the signal in
directions away from the Galactic center are relatively more important.

We find that it takes, in general, shorter amount of time for the
upward muon events than the contained muon ones to reach the desired detection significance. It is mostly because of the
small effective volume of the detector for contained events
which results in lower event rates. The background is also
small in this situation, however, still longer observation times are required due to
low statistics.
Among all the decaying DM
models that we consider, the leptophilic model ($\chi\rightarrow \mu^+\mu^-$) seems most
promising
for detection of DM signal at 2$\sigma$ level for $\theta_{\rm max} > 20^\circ$
 via upward muon events within a few years of observation.

\subsection{Hadronic and Electromagnetic Showers}

In addition to muons,
the showers (hadronic and/or electromagnetic)
produced in neutrino interactions could be used as
 signals of DM.
 The
 shower flux is given by \cite{Dutta}
\begin{eqnarray}\label{showers}
\frac{d\phi_{sh}}{dE_{sh}}&=&  \int^{E_{max}}_{E_{sh}}dE_\nu\left(\frac{d\phi_\nu}{dE_\nu}\right)\frac{N_A\rho}{2}\times\nonumber\\
&\times&\left(\frac{d\sigma^p_\nu(E_\nu,E_\nu-E_{sh})}{dE_{sh}}+(p\rightarrow{n})\right)\nonumber\\
&+& (\nu\rightarrow \bar{\nu}),
\end{eqnarray}
where
${d\sigma^{p,n}}/{dE_{sh}}$ is the differential
cross section for showers produced in neutrino and antineutrino
charged-current and
neutral-current interactions with protons and neutrons.

\begin{figure}[h]
\begin{center}
\epsfig{file=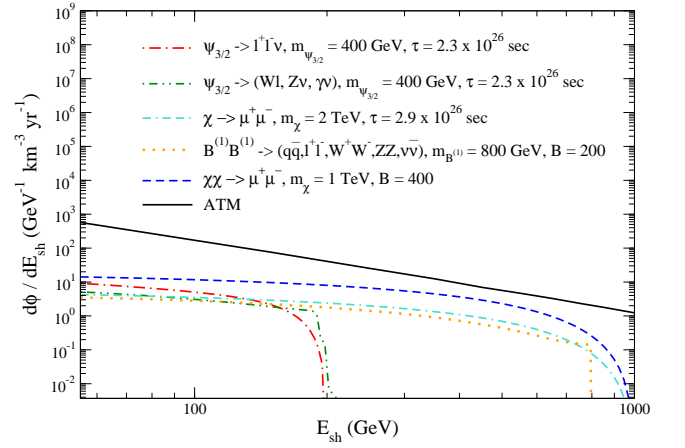,width=2.9in,angle=270}
\end{center}
\caption{Flux for the shower events for two-body (dot-dot-dashed line)
and three-body (dot-dashed line) gravitino decay, decaying leptophilic model (dash-dash-dotted line),
Kaluza-Klein annihilation (dotted line) and annihilating leptophilic model (dashed line) compared with the atmospheric
background (solid line), for the case when
 $\theta_{{\rm max}}=50^\circ$. Model parameters are given in Table II.}
\label{fig:shower_dist}
\end{figure}

In Fig. \ref{fig:shower_dist}.
we show
 shower flux for the case when the DM particle is
a Kaluza-Klein particle, a gravitino and for a leptophilic model, for the
model parameters listed in Table II.  We take
 the
cone half angle around the Galactic center to be $\theta_{\rm max}=50^\circ$.
The shapes of the shower fluxes are similar to the contained muon fluxes
presented in Fig. 4.  The
``golden channel'', $\chi\chi\rightarrow\nu\overline\nu$, as in the case of
contained muons, contributes to the
shower events
at the highest kinematically allowed energy
 for the case when the DM particle is a
Kaluza-Klein particle (dotted line) and when it is a gravitino (dot-dot-dashed line).
While for the case of contained muons (Fig. 4), DM signal in models with  
annihilating DM (Kaluza-Klein and leptophilic) exceeded the atmospheric
background at high energies, shower signals for the same models are below the background.
In addition, shower signals
for the models in which DM annihilate
 are comparable to those in which DM decays. This is mainly due to our
choice of large
$\theta_{\rm max}$ for the showers
($\theta_{\rm max}=50^\circ$).  This can be seen by comparing
$\langle J_2\rangle_\Omega$ (for DM annihilation)
and $\langle J_1\rangle_\Omega$ (for DM decay)
given in
Appendix A (Eq. (\ref{J_factor_n})) for different values of the
cone size.
For example, $\langle J_2\rangle_\Omega/\langle J_1\rangle_\Omega = 75$ for $\theta_{\rm max}=1^\circ$
whereas $\langle J_2\rangle_\Omega/\langle J_1\rangle_\Omega = 3$ for
$\theta_{\rm max}=50^\circ$.
Furthermore, shower fluxes extend to higher energies for the models in
which DM annihilates than the case of DM decay, due to the kinematic constraints.

\begin{figure}[h]
\begin{center}
\epsfig{file=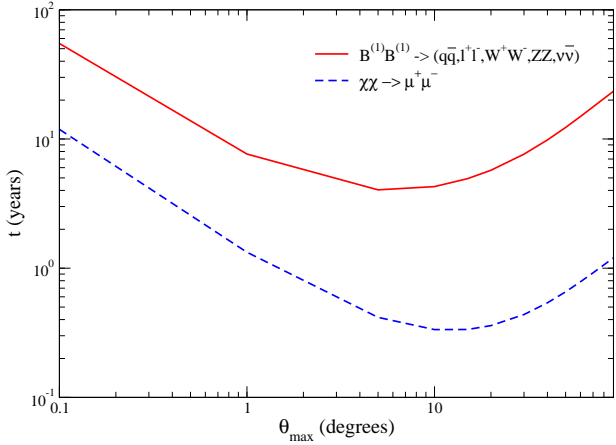,width=2.9in,angle=270}
\end{center}
\caption{
The time $t$ versus cone half angle $\theta_{\rm max}$ centered at the
Galactic center to reach
2$\sigma$ detection significance for the hadronic showers produced
by the neutrinos from $\chi\chi\rightarrow\mu^+\mu^-$ (dashed line) and from 
the Kaluza-Klein DM particle (solid line). We take $E^{th}_{sh}=50$ GeV.
}
\label{fig:shower_sigma}
\end{figure}

\begin{figure}[h]
\begin{center}
\epsfig{file=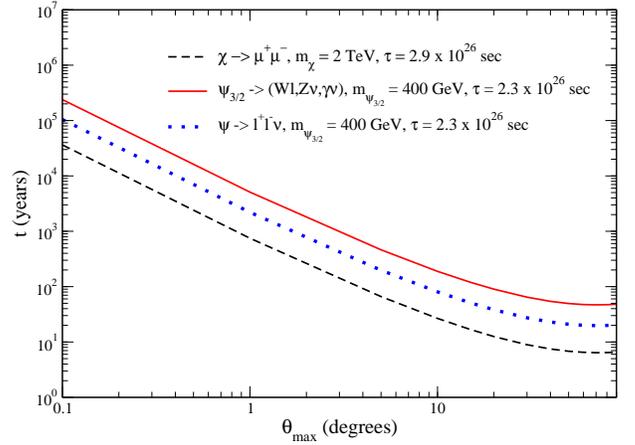,width=2.9in,angle=270}
\end{center}
\caption{
The time $t$ versus cone half angle $\theta_{\rm max}$ centered at the
Galactic center to reach
2$\sigma$ detection significance for the hadronic showers produced
by the neutrinos from $\chi\rightarrow\mu^+\mu^-$ (dashed line) and from the two-body (solid line)
and three-body (dotted line) decay models of gravitino. We take $E^{th}_{sh}=50$ GeV.
}
\label{fig:shower_sigma2}
\end{figure}

We evaluated the shower rates by integrating Eq. (\ref{showers}) over the shower energies
\begin{equation}\label{shower_event}
N_{sh}(m_\chi)=\int^{E_{max}}_{E^{th}_{sh}} \frac{d\phi_{sh}}{dE_{sh}} V_{\rm eff} (E_{sh})
dE_{sh},
\end{equation}   

where
$V_{\rm eff} (E_{sh})$ is the
effective volume of the detector for measuring showers.

In Table \ref{table:shower_rates}, we present the shower rates for a cone size of
$\theta_{\rm max}=50^\circ$ for the DM models shown in Fig. \ref{fig:shower_dist}.
For the detector volume,
we use an energy independent IceCube/DeepCore volume whose value is
$V_{\rm eff}=0.02$ km$^3$ for the showers
\cite{freese2,volume_eff,volume_cascade}.  The detector threshold is
again set to be $E^{th}_{sh}=50$ GeV.  Although the number of events for all
models is relatively small compared to the atmospheric background,
the number of years needed for detection of the $2\sigma$ effect is encouraging
for the leptophilic model as seen in
 Table \ref{table:shower_rates}.
We note that showers seem to be the best way to look for the signal
of decaying DM.

\begin{table}[h]
\centering
\begin{tabular}{l|cc}
\hline\hline
  &\multicolumn{2}{c}{$\theta_{\rm max} = 50^\circ$} \\
\hline
& N$_{sh}$ & t$_{sh}$ (yr)  \\
\hline
$\psi_{3/2}\rightarrow$l$^+$l$^-\nu$ & 11 & 23  \\
$\psi_{3/2}\rightarrow (Wl,Z\nu$,$\gamma\nu$) & 7 & 56 \\
$\chi\rightarrow\mu^+\mu^-$ & 20 & 7  \\
$B^{(1)}B^{(1)}\rightarrow (q\bar{q},l^+l^-,\nu\bar{\nu},...)$ & 15 & 12  \\
$\chi\chi\rightarrow\mu^+\mu^-$ & 68 & 0.64  \\
\hline
ATM & 676 &  \\
\hline
\end{tabular}
\caption{
Shower event rates per year and the time required to reach 2$\sigma$ detection significance for the
 hadronic showers. We consider DM signals from different models
described in the text.  We take the effective volume to be $V_{\rm eff}=0.02 $km$^3$.
}
\label{table:shower_rates}
\end{table}

\begin{figure}[h]
\begin{center}
\epsfig{file=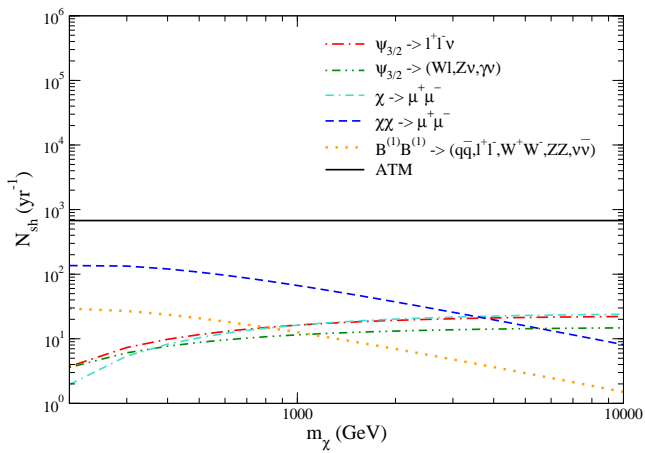,width=2.9in,angle=270}
\end{center}
\caption{Shower event rates as a function of DM mass for models considered in
Table II. The boost factors for  
annihilating DM models and the decay times for the decaying DM models are fixed to the
values in Table II.
We take $\theta_{\rm max}=50^\circ$ and $E^{th}_{sh}=$ 50 GeV. 
}
\label{fig:shower_rates}
\end{figure}

The shower event rates depend on the DM decay time and the boost factor
as an overall normalization while the dependence on DM mass
is non-trivial. In order to illustrate shower event 
rate dependence on the DM mass, we take
fixed boost factor or fixed decay time used in Fig. \ref{fig:shower_dist}
for the annihilating (decaying) DM models and we obtain shower rates for different
 $m_\chi$.
From Fig. \ref{fig:shower_rates} we note
that the shower event rate dependence on $m_\chi$ has similar behavior to
 to the case of contained muon events (see Fig \ref{fig:containedmu_rates}),
due to the similar flux of showers and muons
in neutrino charged-current and neutral-current interactions.
For the case when the DM is a Kaluza-Klein particle or in
the leptophilic ($\chi\chi\rightarrow\mu^+\mu^-$) model the shower rates
decrease with the DM mass while for the decaying DM models it is slowly increasing
 with $m_\chi$ for $m_\chi<1$ TeV and
become almost $m_\chi$ independent for higher values of
$m_\chi$.  For the shower events,
the differential weak scattering cross section in Eq. (\ref{showers})
becomes higher for low shower energies so that the rates for the showers tend to
increase more than those for the muons as the energy decreases.
Consequently, the dependence on the choice of the detector threshold becomes
more significant for the DM masses close to the threshold energy.
We also note that within the cone half angle of
$50^\circ$ and with
the chosen model parameters the rates for showers due to
decaying DM models are comparable to those for the annihilating DM models.
For $m_\chi>1$ TeV ($m_\chi>4$ TeV), the shower rates for decaying DM models are larger than for the 
Kaluza-Klein (leptophilic where $\chi\chi\rightarrow\mu^+\mu^-$) model.

The time required for the shower signal to be at $2\sigma$ level for
different values of the cone half angle centered at the Galactic center is presented
in Fig. \ref{fig:shower_sigma} for the Kaluza-Klein DM particle (solid line)
and for leptophilic model (dashed line) as example for annihilating and decaying DM models.
We find that the optimum time is for the cone half angle of 4$^\circ$ - 10$^\circ$.
For leptophilic model,
one could reach $2 \sigma$ signal within few months of observation
 for $\theta_{max} = 10^\circ$.  In case of Kaluza-Klein DM particle, one needs
about 4 years of data taking for $\theta_{\max}=4^\circ$.
For comparison,
in Fig. \ref{fig:shower_sigma2} we show time needed for detection of $2\sigma$ signal
for the leptophilic model when DM particle decays (dashed line),
gravitino two-body decay (solid line) and gravitino three-body decay (dotted line) models.
The optimum angle for decaying DM models is about
$50^\circ$. This is similar to the previously discussed muon signal presented in
Fig. 13, but shower signals seem to be better for possible detection of the gravitino. This is because of better statistics with the increase in the
cone half angle and also the inclusion of the contributions from
the charged-current and neutral-current interactions of the tau neutrinos
and electron neutrinos with the detector medium. We note, however, that the 20 years  required for
$\theta \geq 50^\circ$ for these parameters is not 
feasible in practice.

\begin{figure}[h]
\begin{center}
\epsfig{file=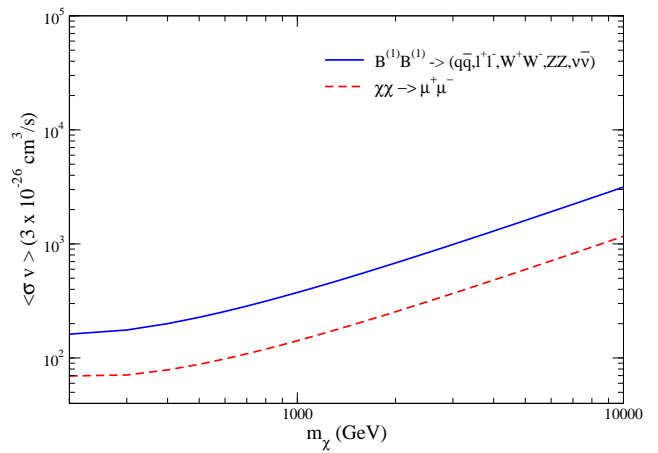,width=2.9in,angle=270}
\end{center}
\caption{
Annihilation cross section versus DM mass for shower
events to reach 2$\sigma$ detection significance after five years of 
observation for the
case of Kaluza-Klein (solid lines) and leptophilic (dashed lines) models.
We take $\theta_{\rm max}=50^\circ$ and $E^{th}_{sh}=$ 50 GeV. 
}
\label{fig:boost_mass3}
\end{figure}

\begin{figure}[h]
\begin{center}
\epsfig{file=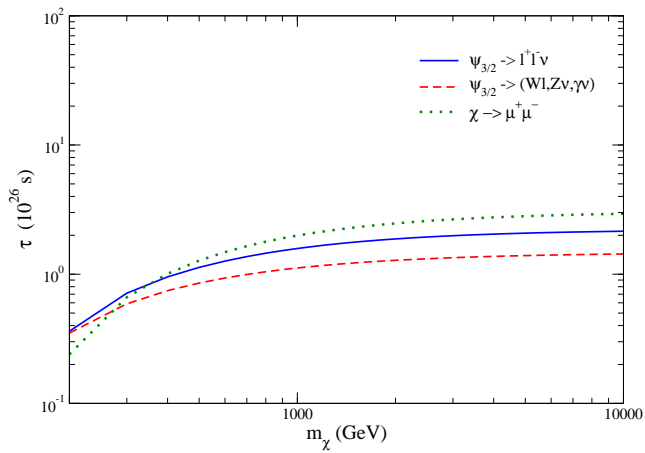,width=2.9in,angle=270}
\end{center}
\caption{ Decay time versus DM mass for both shower
events to reach 2$\sigma$ detection significance after five years of 
observation for the
case of decaying DM models: gravitino three-body(solid lines), gravitino three-body (dashed lines)
and leptophilic (dotted lines). We take $\theta_{\rm max}=50^\circ$ and $E^{th}_{sh}=50$ GeV.
}
\label{fig:decayt_mass3}
\end{figure}

In Fig. \ref{fig:boost_mass3}, we present results for
DM annihilation cross section required for a given
 DM mass
in order to reach 2$\sigma$ detection significance with shower events in 
five years of observation
within $\theta_{\rm max}=1^\circ$.
From Fig. \ref{fig:shower_rates} we note that
 the shower event rates decrease with $m_\chi$ for annihilating DM particles for
fixed annihilation cross section, therefore
the curves for DM annihilation cross section versus DM mass in Fig. \ref{fig:boost_mass3} increase
in order for reach $2\sigma$ effect
for a given $m_\chi$.
The parameter space above the curves defines the exclusion region and the leptophilic model
seems to be more constrained than the Kaluza-Klein model.
For the decaying DM models, our results for shower event rates
 obtained with fixed decay times (Fig. \ref{fig:shower_rates}) indicate that
the 2$\sigma$ detection significance in five years
curves for the
decay time versus DM mass
should increase with $m_\chi$ for $m_\chi<1$ TeV and becomes almost flat for $m_\chi>1$ TeV.
This is shown in Fig. \ref{fig:decayt_mass3}.
In this case, the exclusion regions for the model parameters, decay time and the mass,
at 2$\sigma$ level in five years are
the regions below the
curves.
For the leptophilic models with parameters that satisfy Eq. (2), 
2$\sigma$ signal detection would
imply that the DM particle mass is 250 GeV in case of the annihilating DM,
while for the case of the decaying leptophilic DM, the DM mass would be
3 TeV.

\section{Conclusions}

We have studied neutrino signals
from DM annihilation and decay in the Galactic center assuming NFW profile as the
DM density distribution in the Galaxy.
We have considered models in which DM particle is a
gravitino, a Kaluza-Klein particle and a particle in a leptophilic model.  In case of
the leptophilic model, we have considered both the case of decaying and
annihilating DM.  For a gravitino, we have taken into account both
two-body and three-body decay channels.
For each DM model,
we have calculated contained and upward muons and showers
 using the model parameters that were obtained by
 fitting the excesses in $\gamma$-ray and in the positron or electron plus positron
data from the observations of HESS, PAMELA and FERMI/LAT.
We have used a range of cone half angles 
for the muon and shower events, and we have studied the dependence on
the choice of the cone size and DM mass. 

Our results are summarized in Table \ref{table:summary}. The specific
models which are designed to account for the lepton excesses, listed in Table II,
are indicated by the italic entries (highlighted in red in the online version).
In addition, in Table \ref{table:summary}, we show the event rates and time required
for a $2\sigma$ observation for a range of masses and several values of $\theta_{\rm max}$. Note that 
the event rates for different $\theta_{\rm max}$ can be 
obtained by using rescaling of J-factors,
for example $J_1(50^\circ)/J_1(10^\circ)=9$ and
$J_1(10^\circ)/J_1(1^\circ)=50$ for decaying dark matter models; $J_2(50^\circ)/J_2(10^\circ)=2.7$ and $J_2(10^\circ)/J_2(1^\circ)=7.2$ for
annihilating dark matter models. In Table \ref{table:Jfactor} one can find the 
values of $J(\theta_{\rm max})$.

For the leptophilic model ($\chi\chi\rightarrow\mu^+\mu^-$), for example
with $m_\chi=1$ TeV and $B=400$, the muon flux
due to DM annihilation dominates over muons produced by the
atmospheric neutrinos, for the muon energies
in the range, 200 GeV $< E_\mu < $ 950 GeV (50 GeV$< E_\mu <$750 GeV) for the
contained (upward) muon events with cone half angle $\theta_{\rm max}=1^\circ$.  
The shower event rates 
for this same model
never exceed the atmospheric neutrino induced shower rates, for $\theta_{\rm max}=50^\circ$, 
in part due to the cone size.

For $m_\chi=800$ GeV, $B=200$ and $\theta_{\rm max}=1^\circ$,
the contained muon flux due to annihilating Kaluza-Klein particles
becomes larger than the background for $E_\mu > 300$ GeV and
up to the kinematic cut-off $E_\mu=m_\chi=800$ GeV. The muon flux
is comparable to the background for
 muon energies 200 GeV $< E_\mu <$ 400 GeV
for the upward muons.

With the model parameters that we considered,
the decaying DM models do not produce the muon signal (upward or contained)
that
is above the atmospheric background.
The muon flux from the decaying DM would be
comparable to those for the annihilating DM models only
if the decay times were of the order $\sim$ 10$^{25}$ sec and $\sim 10^{24}$ sec
for
contained muon and upward muon events, respectively.
In contrast to the muon case,
we find that for a wide range of shower energies, the shower 
flux for decaying leptophilic particle
($\chi\rightarrow\mu^+\mu^-$) is larger than for annihilating Kaluza-Klein particle.

We have also calculated the
total muon and shower event rates by folding the corresponding fluxes
 with the energy independent IceCube/DeepCore
effective volume, i.e $V_{\rm eff}=0.04 (0.02)$ km$^3$ for
the contained muon (shower) events. Due to its location IceCube detector is
unable to study the upward muons produced by the neutrinos coming from the Galactic center.
However, we have
 calculated
the upward muon rates
for the IceCube-type detector in the Northern hemisphere
by folding the muon fluxes with
a muon effective area, which is assumed to be
$A_{\rm eff}=1$ km$^2$.

Even if there is a significant signal to background ratio,
low statistics may yield difficulties in confirming the presence of a DM signal
via neutrinos. Thus, we have evaluated how many years it would
take to observe $2\sigma$ effect.
 Using our results for the muon and shower event rates,
we have also obtained the contour exclusion plots in which we show the regions
for the model parameter space for each DM model in the case of
no signal detection at 2$\sigma$ detection significance
in five years.

We find that the leptophilic model ($\chi\chi\rightarrow\mu^+\mu^-$)
has stronger constraints on DM
 annihilation cross section (or the boost factor) and $m_\chi$ than the case of
the
Kaluza-Klein particle. In terms of the constraint on the annihilation cross section,
for the leptophilic DM model where the boost factor and DM mass are related by
Eq. (2), after five years, the range of $m_\chi>250$ GeV would be excluded 
by
upward muon events for $A_{\rm eff}=1$ km$^2$ and events within a cone 
half angle of $\theta_{\rm max}=1^\circ$.
A similar limit is obtained from the shower rate as well.
More generally, we find that
 the upward muon and the shower events are more
constraining than the contained muon ones.
If there is no upward muon signal detected in five years, with 
$\theta_{\rm max}=10^\circ$,
for the decaying leptophilic model ($\chi\rightarrow\mu^+\mu^-$) which satisfies
the constraint given by Eq. (2),
$m_\chi$ is constrained to be smaller than $3$ TeV.

In our calculations, we have taken the detector muon and shower energy thresholds to be 50 GeV.
Changing the detector threshold energy to about 100 GeV does not affect our results significantly.
However, decreasing the threshold energy down to $\sim$ 10 GeV results in a larger
atmospheric background relative to
the DM signal.
Consequently, detecting a DM signal via muon or shower events with
low detector thresholds ($\sim 10$ GeV) becomes more difficult.

Increasing the cone half angle, $\theta_{\rm max}$ about the Galactic center
 increases the DM signal, however it does not necessarily
improve the detection significance since the
background signal due to atmospheric neutrinos is also enhanced.  We have found
 the optimum cone half angles for all types of events in order to
reach 2$\sigma$ detection level. For the annihilating DM models,
we have found the optimum angle to be a few degree for the muon events and about 10$^\circ$
for the shower events. In both cases, we have shown that there is a good chance of detecting both
leptophilic and Kaluza-Klein particles in less than ten years for some DM masses.

In the case of the decaying DM models, the optimum angle is about
50$^\circ$ for both muons and showers.   For gravitino
DM, signals
could be detected in ten years
only with shower events, while the decaying leptophilic particle can be detected
with upward muon events as well in few years.

Our results with fixed annihilation cross section or
decay time and variable DM mass can be used to
predict neutrino signals for any
other values for the parameters of the models we consider.
For example, in the literature, the
HESS data is also explained by a
hypothetical Kaluza-Klein particle with mass
$m_\chi=10$ TeV \cite{KK_HESS}. The boost factor that is required, in this case,
is $B=1000$.  {By rescaling the
results in the previous sections, we find that the observation time to reach 2$\sigma$ detection significance
becomes 0.1 years
for the upward muon events. For contained muons and showers, it is not feasable to detect 2$\sigma$
effect within reasonable time ($t\gg$ 20 years).  
Increasing the boost factor by a factor of 5 and the DM mass by a factor of
about 10 relative to the parameters of Table \ref{table:models}
significantly improves the chance for detecting the Kaluza-Klein particle via
upward muon events.}

Similar to the decaying leptophilic model, the gravitino model ($\psi_{3/2}\rightarrow l^+l^-\nu$)
with a gravitino mass of 3.3 TeV and decay time 
$\tau = 5\times 10^{25}$ sec 
can also
account for the FERMI data \cite{gravitino_PAMELA}. For a decaying DM
particle, increasing the DM mass (for a fixed decay time) increases the event rates.
Since the neutrino flux scales as $\sim\tau^{-1}$ (see
Eq. (\ref{decay_flux})),
decreasing the decay
time also enhances the neutrino signals for a decaying DM particle.
Therefore, combination of
higher mass and shorter decay time should increase all the event rates that
we have calculated for a lighter ($m_{\psi_{3/2}}=400$ GeV)
gravitino particle which has a longer decay time ($\tau=2.3\times 10^{26}$ sec).
For the model parameters, $m_\chi=3.3$ TeV and $\tau = 5\times 10^{25}$ sec 
we find that the 2$\sigma$ detection significance
can be reached in 2.6 years via upward muons and in less than a year via the hadronic showers.
In addition, for the contained muons
the observation time decreases by two orders of magnitude relative to the value given in Table \ref{table:muonevtrates}.


\begin{table*}[p]
\centering
\begin{tabular}{l|c|cccccccccc}
\hline\hline
 & &\multicolumn{9}{c}{$m_\chi$ (TeV)}\\
 & & 0.2 & 0.4 & 0.6 & 0.8 & 1 & 2 & 4 & 6 & 8 & 10 \\
\hline
$\psi_{3/2}\rightarrow l^+ l^-\nu$ & $N_\mu^{ct} (50^\circ)$ & 4.94  & \textcolor{red}{\it 11.15} & 13.8 & 15.3 & 16.2 & 18.1 & 19.0 & 19.3 & 19.5 & 19.6  \\
\textcolor{red}{\it $B_\tau=2.3$} & $N_\mu^{up} (50^\circ)$ & 8.68 & \textcolor{red}{\it 59.5} & 120 & 180 & 239 & 503 & 912 & 1228 & 1485 & 1704  \\
& $N_{sh} (50^\circ)$ & 4 & \textcolor{red}{\it 11} & 13 & 15 & 16.3 & 19 & 21 & 22 & 22 & 22   \\
& $t_\mu^{up} (10^\circ)$ & $1.3\times 10^4$ & \textcolor{red}{\it 277} & 69 & 30 & 17 & 4 & 1.2 & 0.7 & 0.5 & 0.4  \\
& $t_\mu^{up} (50^\circ)$ & 3490 & \textcolor{red}{\it 74} & 18 & 8 & 5 & 1 & 0.32 & 0.18 & 0.12 & 0.09  \\
& $t_{sh} (50^\circ)$ & 196 & \textcolor{red}{\it 23} & 16 & 12 & 10 & 7 & 6.3 & 5.8 & 5.8 & 5.8  \\
\hline
$\psi_{3/2}\rightarrow (Wl,Z\nu,\gamma\nu)$ & $N_\mu^{ct} (50^\circ)$ & 6.1 & \textcolor{red}{\it 8.4} & 8.9 & 9.1 & 9.15 & 9.2 & 9.2 & 9.2 & 9.2 & 9.2  \\
\textcolor{red}{\it $B_\tau=2.3$} & $N_\mu^{up} (50^\circ)$ & 9.9 & \textcolor{red}{\it 50.9} & 95.6 & 139 & 181 & 364 & 638 & 844 & 1010 & 1150  \\
& $N_{sh} (50^\circ)$ & 3.6 & \textcolor{red}{\it 7.66} & 9.6 & 10.74 & 11.5 & 13.17 & 14.12 & 14.46 & 14.64 & 14.74  \\
& $t_\mu^{up} (10^\circ)$ & $1\times 10^4$ & \textcolor{red}{\it 378} & 107 & 51 & 30 & 7.5 & 2.5 & 1.4 & 1 & 0.8  \\
& $t_\mu^{up} (50^\circ)$ & 2693 & \textcolor{red}{\it 101} & 29 & 14 & 8 & 2 & 0.7 & 0.4 & 0.3 & 0.2  \\
& $t_{sh} (50^\circ)$ & 210 & \textcolor{red}{\it 47} & 30 & 24 & 21 & 16 & 14 & 13 & 13 & 13  \\
\hline
$\chi\rightarrow\mu^+\mu^-$ & $N_\mu^{ct} (50^\circ)$ & 2.13 & 6.45 & 8.43 & 9.5 & 10.2 & \textcolor{red}{\it 11.5} & 12.2 & 12.4 & 12.5 & 12.6  \\
\textcolor{red}{\it $B_\tau=2.9$} & $N_\mu^{up} (50^\circ)$ & 3.14 & 29 & 62.3 & 97 & 131 & \textcolor{red}{\it 286} & 533 & 728 & 886 & 1022  \\
& $N_{sh} (50^\circ)$ & 1.95 & 8.22 & 12.09 & 14.55 & 16.2 & \textcolor{red}{\it 20.2} & 22.45 & 23.27 & 23.68 & 23.94  \\
& $t_\mu^{up} (10^\circ)$ & $1\times 10^5$ & $1\times 10^3$ & 252 & 104 & 57 & \textcolor{red}{\it 12} & 3.5 & 1.9 & 1.3 & 0.97  \\
& $t_\mu^{up} (50^\circ)$ & $2.6\times 10^4$ & 316 & 68 & 28 & 15 & \textcolor{red}{\it 3.2} & 0.93 & 0.5 & 0.34 & 0.26  \\
& $t_{sh} (50^\circ)$ & 709 & 40 & 19 & 13 & 11 & \textcolor{red}{\it 6.9} & 5.5 & 5.2 & 5 & 4.8  \\
\hline
$B^{(1)}B^{(1)}\rightarrow ...$ & $N_\mu^{ct} (10^\circ)$ & 14.2 & 9.8 & 7.2 & \textcolor{red}{\it 5.6} & 4.6 & 2.4 & 1.25 & 0.84 & 0.63 & 0.51  \\
\textcolor{red}{\it $B=200$} & $N_\mu^{up} (10^\circ)$ & 86.1 & 131 & 140 & \textcolor{red}{\it 130} & 128 & 124 & 108 & 92 & 81 & 72  \\
& $N_{sh} (10^\circ)$ & 11 & 9 & 7 & \textcolor{red}{\it 5.7} & 4.8 & 2.6 & 1.4 & 0.9 & 0.7 & 0.6  \\
& $t_\mu^{up} (1^\circ)$  & 1.27 & 0.63 & 0.54 & \textcolor{red}{\it 0.65} & 0.66 &0.7 & 0.87 & 1.14 & 1.42 & 1.72  \\
& $t_\mu^{up} (10^\circ)$ & 1.55 & 0.68 & 0.57 & \textcolor{red}{\it 0.71} & 0.72 & 0.76 & 1.0 & 1.36 & 1.76 & 2.2  \\
& $t_\mu^{up} (50^\circ)$ & 5.1 & 2.2 & 1.84 & \textcolor{red}{\it 2.29} & 2.3 & 2.44 & 3.2 & 4.5 & 5.8 & 7.2  \\
& $t_{sh} (1^\circ)$  & 3.4 & 4.4 & 5.9 & \textcolor{red}{\it 7.7} & 9.6 & 22 & 61 & 116 & 189 & 280  \\
& $t_{sh} (10^\circ)$ & 1.3 & 1.9 & 2.9 & \textcolor{red}{\it 4.3} & 5.8 & 18 & 64 & 136 & 237 & 364  \\
& $t_{sh} (50^\circ)$ & 3.3 & 5 & 8 & \textcolor{red}{\it 12} & 16.3 & 57 & 204 & 445 & 777 & 1202  \\
\hline
$\chi\chi\rightarrow\mu^+\mu^-$ & $N_\mu^{ct} (10^\circ)$ & 40.19 & 29.58 & 22.01 & 17.39 & \textcolor{red}{\it 14.3} & 7.59 & 3.90 & 2.63 & 1.98 & 1.59 \\
\textcolor{red}{\it $B=400$} & $N_\mu^{up} (10^\circ)$ & 144 & 241 & 273 & 283 & \textcolor{red}{\it 320} & 266 & 221 & 190 & 167 & 151  \\
& $N_{sh} (10^\circ)$ & 51.4 & 45.6 & 36.4 & 30 & \textcolor{red}{\it 25} & 14 & 7.4 & 5 & 3.8 & 3  \\
& $t_\mu^{ct} (1^\circ)$  & 1.11 & 1.68 & 2.55 & 3.61 & \textcolor{red}{\it 4} & 13.64 & 44 & 92 & 156 & 238  \\
& $t_\mu^{ct} (10^\circ)$ & 0.66 & 1.18 & 2.06 & 3.24 & \textcolor{red}{\it 4.7} & 16.31 & 61 & 133 & 234 & 364  \\  
& $t_\mu^{ct} (50^\circ)$ & 1.93 & 3.55 & 6.38 & 10.2 & \textcolor{red}{\it 15} & 53 & 201 & 444 & 781 & 1213  \\
& $t_\mu^{up} (1^\circ)$  & 0.54 & 0.24 & 0.2 & 0.18 & \textcolor{red}{\it 0.14} & 0.21 & 0.28 & 0.35 & 0.43 & 0.50  \\
& $t_\mu^{up} (10^\circ)$ & 0.47 & 0.21 & 0.16 & 0.15 & \textcolor{red}{\it 0.12} & 0.17 & 0.25 & 0.33 & 0.42 & 0.52  \\
& $t_\mu^{up} (50^\circ)$ & 1.83 & 0.65 & 0.51 & 0.47 & \textcolor{red}{\it 0.37} & 0.54 & 0.78 & 1.1 & 1.35 & 1.7  \\
& $t_{sh} (1^\circ)$  & 0.63  & 0.72 & 0.91 & 1.12 & \textcolor{red}{\it 1.37} & 2.58 & 5.5 & 9 & 13 & 18  \\
& $t_{sh} (10^\circ)$ & 0.12 & 0.14 & 0.2 & 0.26 & \textcolor{red}{\it 0.34} & 0.87 & 2.63 & 5.34 & 9 & 13.6  \\
& $t_{sh} (50^\circ)$ & 0.18 & 0.22 & 0.33 & 0.48 & \textcolor{red}{\it 0.7} & 2.1 & 7.2 & 15.5 & 27 & 42  \\
\hline\hline
Atmospheric & $N_\mu^{ct}$   & \multicolumn{3}{c}{2.28$(1^\circ)$} & \multicolumn{4}{c}{227.5$(10^\circ)$} & \multicolumn{3}{c}{5347$(50^\circ)$} \\
& $N_\mu^{up}$   & \multicolumn{3}{c}{28$(1^\circ)$} & \multicolumn{4}{c}{2794$(10^\circ)$} & \multicolumn{3}{c}{65668$(50^\circ)$} \\
& $N_{sh}$   & \multicolumn{3}{c}{0.3$(1^\circ)$} & \multicolumn{4}{c}{28.8$(10^\circ)$} & \multicolumn{3}{c}{676$(50^\circ)$} \\
\hline\hline
\end{tabular}
\caption{Summary of the results for the event rates and the time that it takes to reach 2$\sigma$ effect for different values of
$m_\chi$ and $\theta_{\rm max}$, where the threshold enery is taken as 50 GeV for both muon and shower events. 
In this table we do not include results which have $t>15$ years for all $m_\chi$. Results for the specific choice of the 
parameters in each model corresponding to fitting PAMELA, FERMI/LAT and HESS data are presented in red italic fonts.
}
\label{table:summary}
\end{table*}


The dependence of the signal on the cone half angle is
different for the annihilating DM particles than for the decaying DM particles.
We have demonstrated this by choosing wedges between $\theta_{\rm max}-1^\circ$ and $\theta_{\rm max}$
centered at the Galactic center and calculating $\langle J_n\rangle_\Omega\Delta\Omega$, which
defines an overall normalization for the
neutrino signals,
for different
$\theta_{\rm max}$ for both annihilating and decaying DM particles.
The angular wedges can be used to rescale event rates as well.
Our results indicate that the $J$ factor
for annihilating DM particle decreases sharply with $\theta_{\rm max}$ whereas for the
case of decaying DM particle, for a wide range of
 $\theta_{\rm max}$,
the $J$ factor has a weak
 $\theta_{\rm max}$ dependence.
In determining the nature of the DM (annihilating or decaying),
the directional dependence of the
neutrino signals gives valuable information.

\acknowledgments{MHR thanks the members of the Center 
for Cosmology and Astroparticle Physics  the Ohio State University 
for their hospitality. This research was supported by 
U.S. Department of Energy Contracts DE-FG02-91ER40664, DE-FG02-04ER41319
and DE-FG02-04ER41298. AEE would like to acknowledge support from 
the University of Arizona Galileo Circle Scholarship.}

\appendix

\section{Neutrino Flux}

The neutrino flux observed at the Earth due to the DM annihilation or decay
 in the galaxy
for a given neutrino flavor can be written as \cite{hisano}

\begin{equation}\label{annihilation_flux}
\frac{d\phi_\nu}{dE_\nu}=R_o\rho^2_o B\frac{\langle\sigma v\rangle}{8\pi m^2_\chi}\left(\sum_F B_F\frac{dN_\nu^{F}}{dE_\nu}\right){\langle J_2
\rangle_\Omega}\Delta\Omega
\end{equation}
for the case of annihilating DM, and

\begin{equation}\label{decay_flux}
\frac{d\phi_\nu}{dE_\nu}=R_o\rho_o \frac{1}{4\pi m_\chi\tau}\left(\sum_F B_F\frac{dN_\nu^{F}}{dE_\nu}\right){\langle J_1
\rangle_\Omega}\Delta\Omega
\end{equation}
for the case of decaying DM where $dN_{\nu}^F/dE_\nu$ is the neutrino spectrum for a given
annihilation or decay channel $F$
with branching fraction $B_F$,
$B$ is the boost factor, $\tau$ is the decay time, $R_o$ is the distance of the solar system from the Galactic center
and $\rho_o$ is the
local density near the solar system.
The neutrino energy spectrum,
$dN_{\nu}^F/dE_\nu$ for different channels can be found in
Appendix B. In the above equations,
the dimensionless
quantity $\langle J_n\rangle_\Omega$ is defined as
\cite{gamma_simul,hisano,liu}

\begin{equation}\label{J_factor_n}
\langle J_n \rangle_\Omega=\int \frac{d\Omega}{\Delta\Omega}\int_{l.o.s}\frac{dl(\theta)}{R_o}\left(\frac{\rho(l)}{\rho_o}\right)^n \; ,
\end{equation}
where $\rho(l)$ is
 the DM density, $l(\theta)$ is the distance between the source and the Earth in
the direction of $\theta$ which is the cone half angle from the Galactic center and the
integral is
 over the
line of sight (l.o.s) within a solid angle $\Delta\Omega = 2\pi(1-cos\theta_{max})$,
centered in the Galactic center.

In our calculations, we take the DM annihilation cross section to have the typical thermal relic value
$\langle \sigma v\rangle=3\times10^{-26}\ {\rm cm}^3
{\rm s}^{-1}$ and we use
the Navarro-Frenk-White (NFW) DM density profile \cite{NFW}

\begin{equation}
\rho(r)=\frac{\rho_s}{(r/R_s)(1+r/R_s)^2} \;\; ,
\end{equation}
where $\rho_s$ and $R_s$ are the parameters which vary from halo to halo.
In our calculations we set $\rho_s=0.2589$ GeV/cm$^3$ and $R_s=20$ kpc so that the DM density
in the vicinity of the solar system ($r=R_o=8.5$ kpc) takes the
typical value $\rho(R_o)=0.3$ GeV/cm$^3$ \cite{salucci}.
Using these definitions, we can write $r=\sqrt{R^2_o+l^2-2R_ol\cos\theta}$ and the upper limit for
the $l$ integral in Eq. (\ref{J_factor_n}) can be obtained as $l_{\rm max}=R_o\cos\theta+\sqrt{R^2_s-R^2_o\sin^2\theta}$
for a given $\theta$.
 
For the NFW profile, some values for
$\langle J_2\rangle_\Omega\Delta\Omega$ and $\langle J_1\rangle_\Omega\Delta\Omega$ are summarized in Table \ref{table:Jfactor}.

\begin{table}[h]
\begin{tabular}{c|c|c|c|c|c|c|c|}
\hline
\hline
 & 0.1$^\circ$ & 1$^\circ$ & 5$^\circ$ & 25$^\circ$ & 50$^\circ$ & 70$^\circ$ & 90$^\circ$ \\
\hline
$\langle J_2\rangle_\Omega\Delta\Omega$ & 0.14 & 1.35 & 5.94 & 19.68 & 27.75 & 31.73 & 33.42   \\
$\langle J_1\rangle_\Omega\Delta\Omega$ & 0.00027 & 0.018 & 0.30 & 3.69 & 8.79 & 12.24 & 14.90 \\
\hline
\end{tabular}
\caption{The values of $J$ factors for NFW profile for $\theta_{max}=0.1^\circ,1^\circ,5^\circ,25^\circ,50^\circ,70^\circ,90^\circ$.}
\label{table:Jfactor}
\end{table}

\begin{figure}[h]
\begin{center}
\epsfig{file=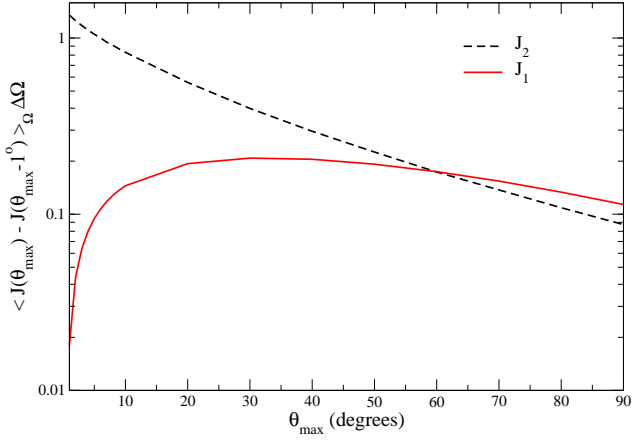,width=2.9in,angle=270}
\end{center}
\caption{$J$ factor values both for annihilating (dashed line) and decaying (solid line) DM models for a wedge between
$\theta_{\rm max}-1^\circ$ and $\theta_{\rm max}$ about the Galactic center as a
function of $\theta_{\rm max}$.}
\label{fig:J_factors}
\end{figure}

In Fig. \ref{fig:J_factors} we show $\langle J_n\rangle_\Omega\Delta\Omega$ 
factors as a function of $\theta_{\rm max}$
for
DM annihilation
($n=2$) and DM decay ($n=1$) evaluated
for a cone wedge between
$\theta_{\rm max}-1^\circ$ and $\theta_{\rm max}$ around the Galactic
center.

\section{Neutrino Spectra}

In this study, we have studied the model dependent
neutrino signals from the annihilation/decay of the DM particles
that reside in the galaxy. We assumed that the DM particles are non-relativistic
so that their total energy is close to their rest energy ($E_\chi\simeq m_\chi$).
The neutrinos with energy $E_\nu$ can be produced from the annihilation/decay of the DM or
from the decay of quarks, charged leptons and gauge bosons which are produced
by the annihilation/decay of the DM.
In our calculations, we have used the standard
unpolarized decay distributions which, in general, take one of the following forms

\begin{equation}\label{3body_same}
\frac{dN}{dx}=2B_f(3x^2-2x^3)   
\end{equation}
\begin{equation}\label{3body_other}
\frac{dN}{dx}=12B_f(x^2-x^3)
\end{equation}
\begin{equation}\label{direct}
\frac{dN}{dx}=B_f\delta(x-1)
\end{equation}
in the rest frame of the decaying particle where $x=2E_\nu/m_d$, $m_d$ is the mass of the
decaying particle and $B_f$ is the decay branching fraction for a given decay
channel. Once the distribution in the rest frame
is known, the neutrino energy spectrum from a decaying particle with velocity
$\beta_d$ and energy $E_d=\gamma_dm_d$ is given by
\cite{boosting_dist}

\begin{equation}\label{boost_equation}  
\left(\frac{dN_\nu}{dE_\nu}\right)=\frac{1}{2}\int_{E_-}^{E_+}\frac{d\epsilon}{\epsilon}\frac{1}{\gamma_d\beta_d}\left(\frac{dN}{d\epsilon}\right)^{\rm
rest} \; ,
\end{equation}
where $E_\mp=E_\nu\gamma_d^{-1}(1\pm\beta_d)^{-1}$.

\subsection{Neutrino spectrum from $\chi\rightarrow Z\nu$ decay channel}

In our calculations, for the channel $\chi\rightarrow{Z}\nu$
we use the Breit-Wigner distribution given as \cite{Covi}

\begin{eqnarray}
\frac{dN_\nu}{dE_\nu} &=& \frac{1}{(E^2_\nu-E^2_{\nu Z})^2+E^2_{\nu Z}\Gamma^2_{\nu Z}}\times\nonumber\\
&\times&\left(\int^\infty_0 \frac{dE}{(E^2-E^2_{\nu Z})^2+E^2_{\nu Z}\Gamma^2_{\nu Z}}\right)^{-1} \; ,
\end{eqnarray}
where the distribution peaks at

\begin{equation}
E_{\nu Z}=\frac{m_\chi}{2}\left(1-\frac{m^2_Z}{m^2_\chi}\right) \; ,
\end{equation}
and  
\begin{equation}
\Gamma_{\nu Z}=\frac{m_Z}{m_\chi}\Gamma_Z .
\end{equation}
We take $\Gamma_Z=2.5$GeV.

\subsection{Neutrino spectrum from $\tau^\mp$, $\mu^\mp$ decay channels}

In this section, we present the $\nu_\mu$ spectrum from $\mu$
and $\tau$ decays. The spectra for other neutrino flavors can be deduced
from these results. For example, the $\nu_e$ spectrum from $\mu$ decay is identical to
the $\nu_\mu$ spectrum from $\tau$ decay and the $\nu_\tau$ spectrum from $\tau$ decay to that
of $\nu_\mu$ from $\mu$ decay. 
For these three-body decays,
the $\nu_\mu$ spectrum from $\tau$ decay is given by Eq. (\ref{3body_other}) and from $\mu$ decay
by Eq. (\ref{3body_same}) in the frame where the decaying particle is at rest.
Then, after boosting these results by using Eq. (\ref{boost_equation})
for charged leptons produced via the
 DM annihilation or decay, we obtain

\begin{equation}\label{neutrino_distribution}
\frac{dN_{\nu_\mu}}{dE_{\nu_\mu}}=\left\{
   \begin{array}{ll}
    \frac{B_f}{E_{l}}(\frac{5}{3}-3x^2+\frac{4}{3}x^3)     &  ,\;\;\mu\rightarrow\nu_\mu e\nu_e \\
    \frac{2B_f}{E_{l}}(1-3x^2+2x^3)     &    ,\;\;\tau\rightarrow\nu_\tau\mu\nu_\mu \; ,
 \end{array}
   \right.
\end{equation}
where $x=\frac{E_{\nu_{\mu}}}{E_{l}}\le 1$ and $E_l=m_\chi$ for the case of annihilation
or $E_l=m_\chi/2$ for the case of decay. The decay branching fraction,
$B_f=0.18 (1)$
for $\tau(\mu)$ decay. In addition to the
three-body decays, $\tau$ can also decay into $\nu_\tau$ via $\tau\rightarrow\nu_\tau M$ or $\tau\rightarrow\nu_\tau X$ where $M=\pi,\rho,a_1$
mesons and $X$ indicates hadrons. The neutrino spectra from these channels are given as \cite{Dutta}

\begin{equation}\label{tau_2body}
\frac{dN_{\nu_\tau}}{dE_{\nu_\tau}}=\left\{
   \begin{array}{lr}
    \frac{B_f}{E_{\tau}}\frac{1}{r_M}\;\;\mbox{when}\;\; \frac{E_\nu}{E_\tau}<r_M     &  \mbox{for mesons} \\
    \frac{0.13}{0.3E_{\tau}}\;\;\mbox{when}\;\;\frac{E_\nu}{E_\tau}<0.3     & \mbox{for hadrons} \; ,
 \end{array}
   \right.
\end{equation}
where $r_M=1-m^2_M/m^2_\tau$ with $m_M$ and $m_\tau$ being the mass of the meson $M$
and the $\tau$ lepton, respectively. Here, $B_f=0.12,0.26,0.13$ and
$r_M=0.99,0.81,0.52$
for $M=\pi,\rho,a_1$ respectively.

\subsection{Neutrino spectrum from
$b(\overline{b})$ and $c(\overline{c})$ decay channels}

The $b$ and $c$ quarks hadronize before they decay into
neutrinos. The hadronization effect is taken into account by scaling
the initial quark energy, $E_{in}$, in the form $E_{f}=z_f E_{in}$,
where $f=b,c$, $z_f=0.73$($0.58$) for $b$ ($c$) quarks
\cite{boosting_dist} and
$E_{in}=m_\chi$ for an annihilating DM particle or $E_{in}=m_\chi/2$ for a decaying DM particle with mass
$m_\chi$.
 
The neutrino spectrum from the decay of
$f$= $b$, $\overline{b}$, $c$ or $\overline{c}$
from $\chi\chi\to f\bar{f}$ can also be approximated by the second equation
 in Eq. (\ref{neutrino_distribution}) (see also
\cite{boosting_dist,lipari}), i.e,

\begin{eqnarray}
\frac{dN_\nu}{dE_\nu}&=&\frac{2B_f}{E_{f}}(1-3x^2+2x^3)\;\;\mbox{for}\;\;x \le 1 \nonumber\\
&=& 0 \;\;\mbox{otherwise} \; ,
\end{eqnarray}
where $x=\frac{E_\nu}{E_{f}}$ and

\begin{equation}
(E_{f}\;,\;B_f)=\left\{
   \begin{array}{lr}
   (0.73E_{in}\;,\;0.103) &       b\;\;  \mbox{channel} \\
   (0.58E_{in}\;,\;0.13)&       c\;\;  \mbox{channel}.
 \end{array}
   \right.
\end{equation}

\subsection{$W^\mp$ and $Z$ decay channels}
 
In the $W^\mp$ and $Z$ decay channels,
the neutrino spectrum from the decaying particle with velocity, $\beta_B$,
and energy, $E_B$,
can be obtained by using Eqs. (\ref{direct}) and (\ref{boost_equation}), i.e,
\begin{eqnarray}
\frac{dN_\nu}{dE_\nu}&=&\frac{n_fB_f}{E_B\beta_B}\;\;\mbox{for}\;\;\frac{E_B}{2}(1-\beta_B)<E_\nu<\frac{E_B}{2}(1+\beta_B)\nonumber\\
&=& 0 \;\;\mbox{otherwise} \; ,
\end{eqnarray}
where

\begin{equation}
(n_f\;,\;B_f)=\left\{
   \begin{array}{lr}
   (1\;,\;0.105) &       W \;\;\mbox{channel}, \\
   (2\;,\;0.067) &       Z\;\;  \mbox{channel}.
   \end{array}
   \right.
\end{equation}

\subsection{Neutrino spectrum from ${t}(\overline{t})$ decay channel}

The top quark decays into a $W$ boson and a $b$ quark ($t\rightarrow{W}{b}$) with a branching fraction
close to unity. Thus, the sum of neutrino spectra of $W$ and $b$ channels gives the required spectrum, i.e,

\begin{equation}
\left(\frac{dN_\nu}{dE_\nu}\right)^{rest}_{t\overline{t}}=\left(\frac{dN_\nu}{dE_\nu}\right)_{W^+W^-}
+
\left(\frac{dN_\nu}{dE_\nu}\right)_{b\overline{b}} .
\end{equation}
Then, boosting this expression yields the neutrino spectrum from top
quarks moving with velocity $\beta_t$ and energy $E_t=\gamma_t m_t$
\cite{boosting_dist},
   
\begin{equation}
\frac{dN_\nu}{dE_\nu}=\left(\frac{dN_\nu}{dE_\nu}\right)_W+\left(\frac{dN_\nu}{dE_\nu}\right)_b \; ,
\end{equation}
where
 
\begin{eqnarray}
\left(\frac{dN_\nu}{dE_\nu}\right)_W &=&
\frac{B_W}{2\gamma_t\beta_tE_W\beta_W}ln\frac{{\rm min}(E_+,\epsilon_+)}{{\rm max}(E_-,\epsilon_-)}\nonumber\\
                  & & \mbox{if}\;\;\gamma_t(1-\beta_t)\epsilon_- < E_\nu <\gamma_t(1+\beta_t)\epsilon_+ \nonumber\\
                  &=& 0 \;\;\mbox{otherwise} \; ,
\end{eqnarray}  
and
 
\begin{eqnarray}
\left(\frac{dN_\nu}{dE_\nu}\right)_b &=&\frac{B_b}{2\gamma_t\beta_tE_d}D_b[E_-/E_d,{\rm min}(1,E_+/E_d)] \nonumber\\
                  & & \mbox{if}\;\;E_\nu < \gamma_t(1+\beta_t)E_d\nonumber\\
                  &=& 0  \;\;\mbox{otherwise} \; ,
\end{eqnarray}
where $B_W=0.105$,$B_b=0.103$,$\epsilon\mp=E_W(1\mp\beta_W)/2$, $E_\mp=E_\nu\gamma_t^{-1}(1\pm\beta_t)^{-1}$
and $E_d=0.73E_b$ with $E_W$,
$\beta_W$ and $E_b$ being
equal to their values in the top-quark rest frame,
i.e,

\begin{eqnarray}
E_b &=&\frac{m^2_t-m^2_W}{2m_t} \nonumber\\
E_W &=& \frac{m^2_t+m^2_W}{2m_t} \nonumber\\
\beta_W &=& \frac{E_b}{E_W}.
\end{eqnarray}
The function    
$D_b$ is given by
\begin{equation}
D_b[m,n]=\frac{1}{3}\left[9(m^2-n^2)-4(m^3-n^3)+6ln\left(\frac{n}{m}\right)\right] .
\end{equation}

\subsection{Neutrino spectrum from $\psi_{3/2}\rightarrow l^+l^-\nu$ decay channel}

In the zero mass limit, the primary lepton ($l^+$ or $l^-$ or $\nu$)
spectrum from the decay channel,
$\psi_{3/2}\rightarrow l^+l^-\nu$,
can be approximated to be

\begin{equation}\label{primary_dist}
\frac{dN_{l(\nu)}}{dE_{l(\nu)}}=\frac{60}{m_{\psi_{3/2}}}x^4(1-x)\;\; \mbox{where} \;\; x=\frac{2E}{m_{\psi_{3/2}}}\le 1 \; ,
\end{equation}
by using the results in \cite{chemtob}. In order to obtain the
spectrum for
the secondary neutrinos produced from the primary charged lepton decays, one can use

\begin{equation}
\frac{dN_\nu}{dE_\nu}=\int_{E_\nu}^{m_{\psi_{3/2}}/2}dE_l\left(\frac{1}{N_l}\frac{dN_l}{dE_l}\right)
\left(\frac{dN_\nu}{dE_\nu}\right)_{l\rightarrow\nu} .
\end{equation}
Here, $\frac{dN_l}{dE_l}$ is the primary charged lepton spectrum given by Eq. (\ref{primary_dist})
and the spectra, $\left(\frac{dN_\nu}{dE_\nu}\right)_{l\rightarrow\nu}$ are
given by Eqs. (\ref{neutrino_distribution}) and (\ref{tau_2body}).
Then, the secondary $\nu_\mu$ spectrum is derived to be

\begin{equation}
\left(\frac{dN_{\nu_\mu}}{dE_{\nu_\mu}}\right)=\frac{5B_f}{m_{\psi_{3/2}}}(1-6x^2+8x^3-3x^4) \; ,
\end{equation}
from the primary $\mu$ decays and   

\begin{equation}
\left(\frac{dN_{\nu_\mu}}{dE_{\nu_\mu}}\right)=\frac{6B_f}{m_{\psi_{3/2}}}(1-10x^2+20x^3-15x^4+4x^5) \; ,
\end{equation}
from the primary $\tau$ decays, requiring that $x\le 1$ in each case
where $x=2E_\nu/m_{\psi_{3/2}}$. Finally, for the $\nu_\tau$ spectrum from the primary $\tau$ decays
accompanied with the meson/hadron production, we find

\begin{equation}
\left(\frac{dN_{\nu_\tau}}{dE_{\nu_\tau}}\right) = \frac{3B_f}{r_Mm_{\psi_{3/2}}}\left(1-\frac{5}{r^4_M}x^4+\frac{4}{r^5_M}x^5\right) \; ,
\end{equation}
with the requirement $r_M>x$.

\section{Parametrizations}

\subsection{Contained and Upward Muons}

Our results for the contained muon fluxes given in Fig. 4 can be parametrized as
\begin{eqnarray}\label{fit:contained}
\frac{d\phi^{ct}_\mu}{dE_\mu} &=& \left(\frac{B}{100}\right)\left(\frac{\langle
J_2\rangle\Delta\Omega}{1.35}\right)
\frac{\xi(x)}{(m_\chi/{\rm TeV})^2}\nonumber\\
\frac{d\phi^{ct}_\mu}{dE_\mu} &=& \left(\frac{\tau}{10^{26}{\rm sec}}\right)^{-1}\left(\frac{\langle
J_1\rangle\Delta\Omega}{0.018}\right)\frac{\xi(x)}{(m_\chi/{\rm TeV})}
\end{eqnarray}  
for annihilation and decay processes, respectively.
The upward muon fluxes presented in Fig. 6 can be parametrized as
\begin{eqnarray}\label{fit:upward}
\frac{d\phi^{up}_\mu}{dE_\mu} &=& \left(\frac{B}{100}\right)\left(\frac{\langle
J_2\rangle\Delta\Omega}{1.35}\right)
\frac{1}{(m_\chi/{\rm TeV})}\frac{C(x)\xi(x)}{(1+1.5\frac{E_\mu}{{\rm TeV}})}
\nonumber\\
\frac{d\phi^{up}_\mu}{dE_\mu} &=& \left(\frac{\tau}{10^{26}{\rm sec}}\right)^{-1}\left(\frac{\langle
J_1\rangle\Delta\Omega}{0.018}\right)
\frac{C(x)\xi(x)}{(1+1.5\frac{E_\mu}{{\rm TeV}})}, 
\end{eqnarray}
where $x=E_\mu/m_\chi$ for DM annihilation and
 $x= 2 E_\mu/m_\chi$ for the DM decay.
The functions $\xi(x)$ and $C(x)$ are fitted to our results for muon fluxes and
parametrizes as
\begin{eqnarray}\label{fit_functions}
\xi(x) &=& a_1+a_2e^{-x}+a_3x+a_4x^2+a_5x^3+\nonumber\\
& &+a_6x^4+a_7\ln(x)\nonumber\\
C(x) &=& 1+a_8x\ ,
\end{eqnarray}
with the best fit values for the parameters given in Tables
(\ref{contained:parameters}) and (\ref{upward:parameters})
for the contained and upward muons, respectively.

\begin{table}[h]
\centering
\begin{tabular}{l|c|c|l|l|l}
\hline\hline  
& $\psi_{3/2}$ & $\psi_{3/2}$ & $\chi$ & $B^{(1)}B^{(1)}$ & $\chi\chi$  \\
& (three-body) & (two-body) & $\rightarrow\mu^+\mu^-$ & $\rightarrow\cdots$ & $\rightarrow\mu^+\mu^-$  \\
\hline
$a_1$ & 16.651 & -32.032  & 0.0554 & 2.5637 & 0.029 \\
$a_2$ & -16.642 & 32.04 & -0.0472 & -2.5470 & -0.0012 \\
$a_3$ & -16.640 & 32.03 & -0.0472 & -2.5546 & -0.0012 \\
$a_4$ & 8.242 & -15.93 & -0.00934 & 1.2381 & -0.11 \\
$a_5$ & -2.553 & 5.022 & 0.0254 & -0.3780 & 0.11 \\
$a_6$ & 0.423 & -0.87 & -0.0069 & 0.06929 & -0.027 \\
$a_7(10^{-6})$ & 0 & 0 & -0.0765 & 215.8 & -0.04 \\
\hline
\end{tabular}
\caption{The best fit parameter values for the contained muon fluxes for different DM models.}
\label{contained:parameters}
\end{table}

\begin{table}[h]
\centering    
\begin{tabular}{l|l|l|l|l|l}
\hline\hline
& $\psi_{3/2}$ & $\psi_{3/2}$ & $\chi$ & $B^{(1)}B^{(1)}$ & $\chi\chi$  \\
& (three-body) & (two-body) & $\rightarrow\mu^+\mu^-$ & $\rightarrow\cdots$  & $\rightarrow\mu^+\mu^-$  \\
\hline
$a_1$ & 1.251 & 1.292 & -0.7716 & -3.4381 & -4.60 \\
$a_2$ & -1.239 & -1.285 & 0.781 & 3.4847 & 4.67 \\   
$a_3$ & -1.266 & -1.296 & 0.747 & 3.3390 & 4.44 \\ 
$a_4$ & 0.638 & 0.641 & -0.341 & -1.5573 & -2.0 \\   
$a_5$ & -0.203 & -0.196 & 0.0918 & 0.4439 & 0.526 \\
$a_6$ & 0.0364 & 0.032 & -0.0132 & -0.0694 & -0.073 \\
$a_7(10^{-6})$ & 0 & 0 & 0.45 & 41.47 & 3.6 \\
$a_8$ & 0.3 & 0.3 & 1.5 & 1.2 & 1.5 \\
\hline
\end{tabular}
\caption{Same as Table (\ref{contained:parameters}) but for the upward muon fluxes.}
\label{upward:parameters}
\end{table}

The background muon flux from atmospheric neutrinos
can be written in a parametric form as

\begin{equation}
\left(\frac{d\phi_\mu}{dE_\mu}\right)^{{\rm ct}}_{{\rm ATM,avg}}=3186.2\left(\frac{E_\mu}{{\rm
GeV}}\right)^{-2.062}\left(\frac{\Delta\Omega}{10^{-3}{\rm sr}}\right)
\end{equation}
in units of GeV$^{-1}$km$^{-3}$yr$^{-1}$ for the contained muon events and

\begin{equation}
\left(\frac{d\phi_\mu}{dE_\mu}\right)^{{\rm up}}_{{\rm ATM,avg}}=89.0\left(\frac{E_\mu}{{\rm GeV}}\right)^{-1.475}
\left(\frac{\Delta\Omega}{10^{-3}{\rm sr}}\right)
\end{equation}
in units of GeV$^{-1}$km$^{-2}$yr$^{-1}$
for the upward muon events.

\subsection{Showers}

Our results for shower flux presented in Fig. 14 can be parametrized as
\begin{eqnarray}\label{fit:showers}
\frac{d\phi_{sh}}{dE_{sh}} &=& \left(\frac{B}{100}\right)\left(\frac{\langle J_2\rangle\Delta\Omega}{27.75}\right)
\frac{\xi(x)}{(m_\chi/{\rm 100 GeV})^2}\nonumber\\
\frac{d\phi_{sh}}{dE_{sh}} &=& \left(\frac{\tau}{10^{26}{\rm sec}}\right)^{-1}\left(\frac{\langle
J_1\rangle\Delta\Omega}{8.79}\right)\frac{\xi(x)}{(m_\chi/{\rm GeV})}
\end{eqnarray}
for DM annihilation and DM
decay processes, respectively.

\begin{table}[h]
\centering
\begin{tabular}{l|c|c|l|l|l}
\hline\hline
& $\psi_{3/2}$ & $\psi_{3/2}$ & $\chi$ & $B^{(1)}B^{(1)}$ & $\chi\chi$  \\
& (three-body) & (two-body) & $\rightarrow\mu^+\mu^-$ & $\rightarrow\cdots$ & $\rightarrow\mu^+\mu^-$  \\
\hline
$a_1$ & -6.335 & -14.129  & -22.867 & -8.765 & -32.43 \\
$a_2$ & 6.355 & 14.139 & 22.898 & 8.78 & 32.48 \\
$a_3$ & 6.274 & 14.100 & 22.765 & 8.72 & 32.29 \\
$a_4$ & -2.991 & -6.952 & -11.134 & -4.25 & -15.79 \\
$a_5$ & 0.813 & 2.127 & 3.325 & 1.25 & 4.71 \\
$a_6$ & -0.099 & -0.347 & -0.5125 & -0.189 & -0.73 \\
$a_7(10^{-3})$ & -0.472 & 0 & 0 & 0 & 0 \\
\hline
\end{tabular} 
\caption{The best fit parameter values for the shower fluxes for different DM models.}
\label{shower:parameters}
\end{table}

The background shower flux from atmospheric neutrinos
can be written in a parametric form as

\begin{equation}
\left(\frac{d\phi_{sh}}{dE_{sh}}\right)_{{\rm ATM,avg}}=3.21\times 10^{6}\left(\frac{E_\mu}{{\rm GeV}}\right)^{-2.155}
\left(\frac{\Delta\Omega}{2.24{\rm sr}}\right)
\end{equation}
in units of GeV$^{-1}$km$^{-3}$yr$^{-1}$.

\end{document}